\documentstyle[12pt]{article}
\renewcommand{\thebibliography}[1]{\section{References}\list
 {[\arabic{enumi}]}{\settowidth\labelwidth{[#1]}\leftmargin\labelwidth
 \advance\leftmargin\labelsep
 \usecounter{enumi}}
 \def\newblock{\hskip .11em plus .33em minus .07em}
 \sloppy\clubpenalty4000\widowpenalty4000
 \sfcode`\.=1000\relax}
\renewcommand{\abstract}{\if@twocolumn
 \section*{Abstract}
 \else \small

\begin{center}
{\bf ABSTRACT\vspace{-.5em}\vspace{0pt}} 

\end{center}

 \quotation 
 \fi}

\title{Anomalous transport: a mathematical framework
\thanks{to appear in Reviews in Mathematical Physics}}

\author{H. Schulz-Baldes\thanks {e-mail: hermann@irsamc2.ups-tlse.fr} ,  
J. Bellissard\thanks {e-mail: jeanbel@irsamc2.ups-tlse.fr}\\
Universit\'e Paul-Sabatier, Toulouse, France \thanks {UMR 5626, CNRS 
and Laboratoire de Physique Quantique, 
118, Route de Narbonne, 31062-Toulouse Cedex, France}}

\date{ }

\newtheorem{theo}{Theorem}
\newtheorem{defini}{Definition}
\newtheorem{proposi}{Proposition}
\newtheorem{lemma}{Lemma}
\newtheorem{coro}{Corollary}
\newtheorem{rem}{Remark}

\newcommand{\CS}{$C^{\ast}$-algebra }

\newcommand{\be}{\begin{eqnarray}}
\newcommand{\ee}{\end{eqnarray}}
\newcommand{\beqn}{\begin{eqnarray}}
\newcommand{\eeqn}{\end{eqnarray}}
\newcommand{\bed}{\begin{description}}
\newcommand{\ed}{\end{description}}

\newcommand{\TV}{{\cal T}}

\newcommand{\Aa}{{\cal A}}
\newcommand{\Bb}{{\cal B}}
\newcommand{\Cc}{{\cal C}}
\newcommand{\Dd}{{\cal D}}

\newcommand{\Ff}{{\cal F}}
\newcommand{\Gg}{{\cal G}}
\newcommand{\Hh}{{\cal H}}
\newcommand{\hH}{{\hat{H}}}
\newcommand{\hHV}{{{\hat{H}}+{\hat{V}}}}
\newcommand{\HV}{{{\hat{H}},{\hat{V}}}}

\newcommand{\Ll}{{\cal L}}
\newcommand{\Nn}{{\cal N}}
\newcommand{\Mm}{{\cal M}}
\newcommand{\Uu}{{\cal U}}
\newcommand{\MH}{{\hat{\cal M}}}
\newcommand{\Oo}{{\cal O}}

\newcommand{\Rr}{{\cal R}}
\newcommand{\Ss}{{\cal S}}

\newcommand{\BB}{{\bf B}}
\newcommand{\CC}{{\bf C}}
\newcommand{\EE}{{\bf E}}
\newcommand{\HH}{{\cal H}}
\newcommand{\NN}{{\bf N}}

\newcommand{\PP}{{\bf P}}

\newcommand{\RR}{{\bf R}}

\newcommand{\ZZ}{{\bf Z}}

\newcommand{\XV}{\vec{X}}

\newcommand{\trel}{\tau_{\mbox{\tiny rel}}}

\newcommand{\au}{\alpha}

\def\essinf{\mathop{\rm essinf}}
\def\esssup{\mathop{\rm esssup}}

\begin{document}

\maketitle

\begin{abstract}

We develop a mathematical framework allowing to study anomalous transport in
homogeneous solids. The main tools characterizing the anomalous transport
properties are spectral and diffusion exponents associated to the covariant
Hamiltonians describing these media. The diffusion exponents characterize the
spectral measures entering in Kubo's formula for the conductivity and hence lead
to anomalies in Drude's formula. We give several formulas allowing to calculate
these exponents and treat, as an example, Wegner's $n$-orbital model as well as
the Anderson model in coherent potential approximation. 

\end{abstract}


\vspace{.5cm}

\section{Introduction}
\label{chap-intro}
\subsection{Anomalous electronic transport}
\label{sec-anomtrans}

Quantum effects and interactions in various materials cause a great
variety of behaviors for electronic transport at low temperature.
Understanding why some materials are conductors and others insulators
is a challenging central problem of solid state physics. The first
attempt to get a microscopic theory of electronic transport goes back
to the work of Drude \cite{Dru} who wrote the conductivity in the form:

\begin{equation}
\label{eq-drudeformula}
\sigma \;= \;\frac{nq^2\tau}{m_{\ast}}
\mbox{ , }
\end{equation}

\noindent where $n$ is the charge carrier density, $q$ is the carrier
electric charge, $m_{\ast}$ is the carrier effective mass and $\tau$ is
the collision time. The derivation of this formula was initially given
in terms of kinetic transport of classical particles, but Sommerfeld
and Peierls rederived it in the context of quantum theory of crystals
which led to consequences in better agreement with experiments in
metallic samples \cite{MA}. The main weakness of the theory lies in the
definition of $\tau$. The collision time is often understood as a
phenomenological parameter that can be easily measured, but is
difficult to interpret. This is the so-called {\em relaxation time
approximation} (RTA). However, using more sophisticated theories, one
can calculate it if various contributions such as electron-impurity,
electron-phonon or electron-electron scattering are taken into account.
It gives a temperature dependence in the form of a power law, namely
$\tau(T)\sim T^{\gamma}$, where $\gamma$ depends upon the type of
collision which dominates dissipation \cite{MA}. 

\vspace{.2cm} 

Among the various mechanisms that may lead to a metal-insulator
transition, one is the {\em anomalous quantum transport}. This means
that, within a one-electron effective theory, the electronic wave
packet diffuses anomalously through the systems instead of moving
ballistically between collisions as  free electrons in a perfect
crystal. Such a mechanism is probably at the basis of the strange
transport properties of quasicrystals at low temperature
\cite{Ber,May}. The best samples are nowadays alloys made of very good
metals, such as the $Al_{62.5}Cu_{25}Fe_{12.5}$,
$Al_{70.5}Pd_{22}Mn_{7.5}$ or $Al_{70.5}Pd_{21}Re_{8.5}$, crystallizing
in a quasiperiodic icosahedral phase. They have conductivities
comparable to doped semiconductors, namely $10^{-6}$ to $10^{-9}$ times
less than for pure aluminum at $4K$ (see for instance
\cite{PPG,RoTrLaiMa} for the following informations); moreover, the
conductivity increases with temperature, which is just the opposite of
the behavior of a normal metal; the temperature dependence of the
conductivity is neither of the exponential type $exp(-E/k_B T)$
characterizing thermally activated processes, nor of the form
$exp(-aT^{-\alpha})$ as in Mott's hopping conductivity or related
mechanisms \cite{ES}, but rather exhibits a power law behavior
$\sigma(T) \sim T^{\beta}$ in the range $4-800K$ with $1 <\beta <1.5$
\cite{RoTrLaiMa}; at last, the behavior of the conductivity as a
function of the magnetic field shows the typical weak localization
signature observed in slightly disordered metals \cite{RoTrLaiMa}.

On the other hand, several numerical results concerning the time
behavior of tight binding models on a quasiperiodic lattice have shown
that the spreading of the wave packet satisfies a power law behavior
in time of the type \cite{HA1,GK,SPB}

\begin{equation}
\label{eq-anomdiff}
\langle \phi |(\XV(t)-\XV(0))^2|\phi\rangle\; \sim
\; t^{2\sigma_{\mbox{\tiny\rm diff}}}
\hspace{1.5cm} 
\mbox{\rm as }
\hspace{.5cm}t\rightarrow +\infty
\mbox{ , }
\end{equation}

\noindent where $|\phi\rangle$ is the initial localized  wave function
and $\XV$ is the position operator for the particle. Depending on the
strength of the quasiperiodic potential, the diffusion exponent
$\sigma_{\mbox{\tiny\rm diff}}$ may vary from $0$ to $0.8$ in a nearest
neighbor model on an octagonal lattice \cite{SPB}. Similar behavior
were  observed in the Harper model \cite{HA1,GK}, the Fibonacci
Hamiltonian \cite{HA1} and the kicked Harper model \cite{ACS}. For
quasicrystals, analytical calculations in one dimension \cite{Pie2} or
more concrete phenomenological models in three dimension \cite{Jan}
have led to similar results and to predictions for the diffusion
exponent $\sigma_{\tiny\rm diff}$.

Experimental results, their theoretical interpretation and numerical
simulations have therefore led the experts toward the idea that
anomalous diffusion as described in equation (\ref{eq-anomdiff}) is the
main reason for the strange transport properties of quasicrystals, at
least for temperatures above approximately $4K$ (below $4K$,
electron-electron interactions may become more important). One
consequence of (\ref{eq-anomdiff}), as we will show in this paper, is
that the conductivity in RTA no longer satisfies the Drude formula
(\ref{eq-drudeformula}), but rather:

\begin{equation}
\label{eq-anomdrude}
\sigma\; \sim\; \tau^{2\sigma_{\tiny\rm diff} -1}
\hspace{1.5cm} 
\mbox{\rm as }
\hspace{.5cm}\tau \rightarrow +\infty
\mbox{ . }
\end{equation}

\noindent Such a formula can be guessed by means of the following non
rigorous argument \cite{SG,May}. By the Einstein relation, $\sigma =
q^2 N(E_F) D(\tau)$ where $N(E_F)$ is the density of states at the
Fermi level $E_F$, while $D(\tau)$ is the diffusion coefficient. One
has $D(\tau)=L(\tau)^2/3\tau$ where $L(\tau)$ is the mean free path,
namely the distance spread by a typical wave packet during the time
$\tau$. Then (\ref{eq-anomdiff}) gives (\ref{eq-anomdrude}).
Consequently, provided $\sigma_{\tiny\rm diff}<1/2$ and $\tau$
depends on temperature as indicated above, the conductivity increases
with temperature. For electrons in a periodic crystal,
$\sigma_{\tiny\rm diff} =1$ (ballistic quantum motion) and equation
(\ref{eq-anomdrude}) gives a weak form of the Drude formula. 

\vspace{.2cm}

Our aim in this paper is to put this type of argument on a rigorous
ground. As we shall see, most of the work consists in defining the
relevant mathematical framework liable to lead to (\ref{eq-anomdrude}).
Then using many recent results on the characterization of singular
continuous spectral measures, we will explain how to justify these
arguments. 

\vspace{.2cm}

\subsection{Homogeneous media}
\label{sec-NCBZ}
 
The first difficulty in dealing with materials such as quasicrystals is
their lack of translation invariance at the atomic scale. In
particular, there is no Bloch theorem allowing analytical calculations.
However, these materials are homogeneous, namely they become
translation invariant at a larger scale. In the limit where the sample
size is large, it becomes simpler to represent these systems in the
{\em infinite volume limit}. This is the first assumption made here. It
means that we exclude the description of mesoscopic devices from our
framework. 

Our next assumption is to work within the {\em one electron
approximation}, that is electron-electron and electron-phonon
interactions are neglected in the Hamiltonian description and treated
only in RTA by means of the collision time. This approximation is
usually very good in metals because electrons can be described as
quasiparticles dressed by interactions. As we have indicated
previously, this is not a restriction when considering quasicrystals as
long as the temperature is not smaller than $4K$. Moreover, this framework
includes electron-impurity scattering in the limit where imputities are
quenched, namely their dynamic is not considered.

The third assumption concerns the energy range within which the
electronic motion is studied. For indeed, only electrons with
energies within $\Oo(k_B T)$ from the Fermi level contribute to the
electronic transport. This allows to describe the electronic motion
by means of an effective one-particle  Hamiltonian, namely the
restriction of the Schr\"odinger operator to this energy interval. In
practice, it is possible to work within the so-called {\em tight
binding representation} (see for instance \cite{Be85}): we consider the
nuclei as fixed on the vertices of the crystalline lattice $\Ll$ and we
restrict ourselves to bands crossing the Fermi level; if only one band
contributes to the Fermi level, wave functions are represented by a
square summable sequence $\psi =(\psi(x))_{x\in \Ll}$, namely an
element of $\Hh =\ell^2 (\Ll)$; the Hamiltonian is then a bounded
selfadjoint operator $H=H^{\ast}$ on $\Hh$ whith off-diagonal matrix
elements $\langle x|H|y\rangle$  decreasing exponentially fast to zero
as $|x-y|\rightarrow \infty$; in practice, one considers only the
nearest neighbors contribution to this operator.  If $M$ bands must be
taken into account, the wave function gets a band index $\psi =(\psi_m
(x)), 1\leq m\leq M$, and $\Hh = \ell^2 (\Ll)\otimes \CC^M$. In this
way we avoid technicalities due to the unboundedness of the Schr\"odinger
operator without restricting the physical domain of applicability. In order
to avoid further irrelevant technical difficulties, we will considered in this
paper systems in which $\Ll$ is a Bravais lattice of the type $\ZZ^d$. In the
case of quasicrystals, this procedure must be slightly modified using groupoids
(see for instance \cite{Be85,Be91,Kel,Con}), but leads to a very similar
description. 

How can now {\em homogeneity} be described? Let us repeat the main
arguments as given by one of the authors \cite{Be85} leading to its
mathematical definition:  since the medium is translation invariant at
large scale, translating $H$ should give the same physical properties.
So there is no reason to prefer $H$ to any other of its translated.
Moreover, changing from sample to sample is equivalent to looking at
one unique infinite system through windows centered at different
locations. This means that if we need to consider limits, $H$ is now by
definition homogeneous if  the strong closure of the set of all
translated, denoted by $\Omega$, is compact. The choice of the strong
topology is crucial: the norm topology leads to the too restricted
notion of almost periodic operators while the weak topology would lead
to a too wide notion of homogeneity. For an unbounded Hamiltonians, a
similar construction holds when the strong resolvent topology is used
\cite{Be91}. $\Omega$ inherits a structure of a topological dynamical
system by considering the action of the translation group on it. This
dynamical system will be called the {\em hull} of $H$.  

This framework will be used to describe robust, namely sample
independent, physical properties. We will focus here on the measurement
of space averages (or equivalently sample averages) of observables. By
{\em observable} we mean any bounded operator obtained from the
Hamiltonian through the elementary operations of translation, sums,
product with a scalar, product, involution and limits in the strong
topology. In this way the {\em observable algebra} contains no more
than the energy and the homogeneity of the system. On the other hand,
space averaging is not uniquely defined in general for its
definition requires the choice of a translation invariant ergodic
probability measure $\PP$ on the hull. This choice will be arbitrary in what
follows because all our results are independent of it. However, one could wonder
whether there is a preferred choice in nature. This is probably related to the
question of existence (and uniqueness) of a Gibbs measure with respect to the
translation dynamic, but we will not address this question here.

\vspace{.2cm}

These considerations lead to the construction of a \CS $\Aa$ that will
be given in Sections \ref{sec-stab} and \ref{sec-Brillouin} below. This
\CS has been called the {\em Non Commutative Brillouin Zone} (NCBZ) of
the homogeneous medium under consideration (see \cite{Be85, Be91}). In
fact, in the periodic case, it coincides with the space of (matrix
valued) continuous functions on the Brillouin zone. The probability
$\PP$ then defines a unique trace $\TV$ on $\Aa$ which is nothing but
the {\em trace per unit volume}. For a periodic crystal, it reduces to
integration over the Brillouin zone. 

\vspace{.2cm}

One advantage of this formalism is that quantities computed through the
trace $\TV$ are insensitive to perturbations of the Hamiltonian by
compact operators (see Theorem \ref{theo-stab} below).  This is
actually a delicate point.  A compact perturbation of the Hamiltonian
can physically be interpreted as a localized impurity in the crystal.
An experimentalist considers relevant only those properties that are
insensitive to a given localized impurity, unless there is a definite
procedure to faithfully reproduce this impurity from sample to sample.
In homogeneous media, such a control is usually out of reach even for
mesoscopic devices.

However, there are mathematical properties depending upon the occurence
of a specific impurity. There have been quite a lot of works, starting
with the result of Donoghue \cite{Don}, showing that a rank one
perturbation of a Hamiltonian with pure-point spectrum may produce a
continuous spectrum (see e.g. \cite{Howl} for a short review of
historical references). One of the most striking results in this
respect has been put in a systematic form in \cite{SW,Go,RJLS1,RJLS2}.
In particular, for the Anderson model in a regime of pure point
spectrum with exponentially localized eigenstates, a topologically
generic rank-one perturbation of the Hamiltonian produces a singular
continuous spectrum \cite{RJLS2}. Moreover, the authors of \cite{RJLS2}
show that the corresponding robust property is that the spectral and
diffusion exponent do not change under such a perturbation and are
equal to zero. This result is mathematically non trivial and
remarkable. Their relevance in practical situation is however
questionable.

In most experimental situations or numerical simulations, physicists
are looking at sample independent quantities such as the average
localization length, the conductance or the magnetoresistance. In more
delicate questions, they may even look at universal fluctuations (a
question not investigated here). In any case, the occurence of only a
volume independent finite number of impurities is out of reach in
practice. This is the main reason why we chose a mathematical framework
in which delicate results, such as the topological instability of
spectral properties by compact perturbations, can be discarded.

\vspace{.2cm}

\subsection{Spectral and transport exponents}
\label{sec-exponents}
 
As we have seen in Section \ref{sec-anomtrans}, scaling laws are the
rule if anomalous transport occurs. This leads us to address the
question of a proper definition of scaling exponents. Several
inequivalent definitions are available in the literature dealing with
fractal and multifractal analysis \cite{Rod,HP,Kad,Fal,Ols}. In this
work, we review several of them and discuss their relevance for our
purpose. More precisely, we shall study the exponents characterizing
the local behavior of the following three Borel measures: (i) the {\em
Density of States} (DOS) expressing global properties related to the
thermodynamics of the electron gas; (ii) the {\em Local Density of
States} (LDOS), namely the spectral measure ``with $\PP$-probability
one''; (iii) the {\em current-current correlation measure} involved in
Kubo's formula for the conductivity.

\vspace{.2cm}

The basic definition of local exponents chosen here follows a
suggestion of G. Mantica (see the second reference in \cite{Gua}).
Given a non-negative Lebesgue measurable function $f$ on $(0,1]$ we say
$ f(\epsilon) \stackrel{{\textstyle \sim}}{{\scriptscriptstyle
\epsilon\downarrow 0}} \epsilon^{\alpha} $ whenever $\int_0^1 d\epsilon
\;\epsilon^{-1-\gamma} f(\epsilon)$ converges for $\gamma<\alpha$ and
diverges for $\gamma>\alpha$. If $f$ is monotonous, this is equivalent
to $\alpha=$liminf$_{\epsilon\rightarrow 0} {\mbox{Log }\; f
(\epsilon)}/{\mbox{Log}\;\epsilon}$. A similar definition holds for the
behavior at infinity. Note that this definition ignores all kinds of
subdominant contributions and is likely to be robust. The local
exponent $\alpha_\nu(E)$ of a Borel measure $\nu$ on $\RR$ is
introduced by 

$$
\int^{E+\epsilon}_{E-\epsilon}d\nu(E') \;\stackrel{{\textstyle
\sim}}{{\scriptscriptstyle  \epsilon\downarrow 0}}
\;\epsilon^{\alpha_\nu(E)}
\mbox{ . }
$$

\noindent We show that $E\mapsto\alpha_\nu(E)$ defines a function in
$L^{\infty}(\RR,d\nu)$ which depends only on the measure class of $\nu$
(see Theorem \ref{theo-expoppac} below). Note that, although a
consequence of standard measure-theoretic arguments, this result cannot
be found in the literature
\cite{Rod,You,Gua,Fal,Cut,Hol,Pes,Com,Ols,Las,RJLS2,BCM}. Note further that
other exponents defined in multifractal analysis do depend upon the
measure in its own equivalence class (see Remark \ref{rem-mesudep} in
Section \ref{sec-global}).

These properties are important in view of applications to homogeneous
systems. A multifractal analysis of the DOS in the vicinity of the
Fermi level may be useful for the thermodynamical properties of
electron gas in our system because it gives more precise information
about the DOS than the local exponents. However, multifractal
properties of the LDOS are of no use in a homogeneous system where only
the measure class of the spectral measure has some robustness. For
indeed, by looking at the system through local windows chosen at random
in the lattice, the corresponding spectral measures are in the same
measure class and the entire equivalence class can be described in this
way (with probability one). Therefore, we cannot expect exponents that do
depend upon the spectral measure in its measure class to be relevant
in practice. 

The local exponents take values in the interval $[0,1]$ $\nu$-almost
surely. For an absolutely continuous measure $\nu$, $\alpha_\nu(E)=1$
$\nu$-almost surely. For a pure point measure $\nu$, $\alpha_\nu(E)=0$
$\nu$-almost surely. Hence these exponents allow to distinguish between
different singular continuous measures. Examples of Hamiltonians with
singular continuous spectra have been studied over years (see
\cite{AS,Si1,BBM,GFP,DP,Cas,Su1,Su2,BIST,Be90,DPY,BBG,BGh,HKS,JS}) and
the question of computing their spectral exponents is certainly worth of
study \cite{JL}. The local exponents can be computed both numerically
and analytically by using the Green's function \cite{ZMB} (see also
\cite{RJLS2}):

\begin{equation}
\label{eq-greenintro}
\Im m\;\int_\RR 
 \frac{d\nu(E')}{E'-(E-\imath\epsilon)}
  \;\stackrel{{\textstyle \sim}}{{\scriptscriptstyle 
   \epsilon\downarrow 0}}  
    \;\epsilon^{\alpha_\nu(E)-1}
\mbox{ . }
\end{equation}

\noindent Note, however, that numerical computations may concern
exceptional values of $E$ for which $\alpha_\nu(E)$ is larger than
$1$.

\vspace{.2cm}

The definition of local exponents extend to the spectral analysis of a
self-adjoint operator on a separable Hilbert space. We show that the
exponents are invariants of the operator itself independent of the
states in the Hilbert space. For covariant families of such
operators arising in homogeneous media as discussed in Section
\ref{sec-NCBZ}, the exponents are moreover $\PP$-almost surely constant
and define the exponents $\alpha_{\mbox{\rm\tiny LDOS}}(E)$ of the
LDOS. The DOS is always regular with respect to the LDOS in the sense
that for typical energies one has $\alpha_{\mbox{\rm\tiny LDOS}}(E)\leq
\alpha_{\mbox{\rm\tiny DOS}}(E)$.

\vspace{.2cm}

The definition of the diffusion exponent $\sigma_{\mbox{\tiny\rm
diff}}$ given in equation~(\ref{eq-anomdiff}) is generalized by
restricting the dynamics to an energy interval $\Delta$ and by
extending it to the case of homogeneous systems. One talks of
ballistic motion whenever $\sigma_{\mbox{\tiny\rm diff}}(\Delta)=1$ and
of regular diffusion whenever $\sigma_{\mbox{\tiny\rm
diff}}(\Delta)=1/2$. Localization, a behavior strictly stronger than
$\sigma_{\mbox{\tiny\rm diff}}(\Delta)=0$, has been studied in \cite{Be94,BES}
and will be discussed in more detail in Section \ref{sec-diffuexpo}. For any
other value of $\sigma_{\mbox{\tiny\rm diff}}$, the quantum diffusion is called
anomalous. Guarneri's inequality \cite{Gua} (see also \cite{Com,MG,Las,BCM})
gives a lower bound of the diffusion exponent by the local exponents of the
LDOS: 

$$
\alpha_{\mbox{\rm\tiny LDOS}}\;\leq\;d\cdot
\sigma_{\mbox{\tiny\rm diff}}
\mbox{ , }
$$

\noindent where $d$ is the space dimension (see Chapter \ref{chap-resu}
for a precise formulation). 

Guarneri's inequality has several direct physical implications
\cite{BeS}. For dimension one, an absolutely continuous spectrum
implies ballistic quantum motion. However, for $d\geq 2$, one may have
both absolutely continuous spectrum and quantum diffusion with
$\sigma_{\mbox{\rm\tiny diff}}\geq 1/d$.  This is expected, in
particular, for the three-dimensional Anderson model at low disorder
where, on the basis of the renormalization group calculation, the
diffusion exponent is conjectured to be $\sigma_{\mbox{\tiny\rm
diff}}=1/2$ \cite{Tho,AALR}. The same situation is expected for the
Anderson model in dimension two provided spin-orbit coupling is added
\cite{HLN}. For three-dimensional quasicrystals, Guarneri's inequality
allows a diffusion exponent as low as $1/3$ without forbidding an
absolutely continuous spectrum. In Chapter \ref{chap-Andfree}, we give
an example of a model having  $\sigma_{\mbox{\tiny\rm diff}}=1/2$. Note
that this model is an operator theoretic version of the so-called {\em
coherent potential approximation} of the Anderson model \cite{NS} .

\subsection{Overview of this article}
\label{sec-organisation}

After this motivating introduction, we have organized this article as
follows. Chapter \ref{chap-resu} contains the main results and
discriminates between the new results and the ones already obtained
elsewhere. In Sections \ref{sec-stab} and \ref{sec-Brillouin} we give
an account of the mathematical framework introduced in \cite{Be85,Be91}
and used to describe homogeneous media. We also give a precise
formulation to the stability under compact perturbations of the
Hamiltonian. Both well-known and new results on local spectral
exponents of Borel measures are presented in Section \ref{sec-expoess}
before being extended to self-adjoint operators in Section
\ref{sec-expoop}. In Sections \ref{sec-DOSLDOS} through \ref{sec-Kubo}
are devoted to the exponents of the LDOS, DOS and current-current
correlation function as well as the anomalous Drude formula.

Chapter \ref{chap-gener} contains proofs of the results of Sections
\ref{sec-expoess} and \ref{sec-expoop} as well as some complementary
results. In particular, in Sections \ref{sec-global} and \ref{sec-CT}
we give few known or less known facts about multifractal analysis which
are of interest.

The remaining results of Chapter \ref{chap-resu}, all linked to
homogeneous systems, are proved in Chapter \ref{chap-expon}. The
content of this chapter has not yet been treated in the literature. 

Chapter \ref{chap-Andfree} is devoted to the calculation of the
diffusion exponent in the Anderson model with free random variables 
\cite{NS}.

An Appendix completes the study of the hull whenever an impurity (or a
compact perturbation) is added to the Hamiltonian.


\vspace{.5cm}

\section{Notations and results}
\label{chap-resu}

\subsection{Construction and stability of the hull}
\label{sec-stab}

Let $\hH=\hH^*$ be a bounded Hamiltonian acting on the one-particle
Hilbert space $\Hh=\ell^2(\ZZ^d)$. Let us consider its hull
$\Omega_\hH$ given by

\begin{equation}
\label{eq-hull}
\Omega_\hH\;=\;\overline{\{U(a)\hH U(a)^{-1}\;|\;a\in\ZZ^d\} 
}^{\;\mbox{\rm\tiny s}}
\mbox{ , }
\end{equation}

\noindent where $(U(a))_{a\in\ZZ^d}$ is a projective unitary
representation of $\ZZ^d$ on $\Hh$ and the closure is taken with
respect to the strong operator topology. By definition $\hH$ is
homogeneous if $\Omega_\hH$ is a compact metrisable space
\cite{Be85,Be91}. The projective representation $U$ induces a
$\ZZ^d$-action $T$ on $\Omega_\hH$ by homeomorphisms. Each point
$\omega\in\Omega_\hH$ describes a disorder or aperiodicity
configuration of the crystal. A $T$-invariant and ergodic probability
measure $\PP$ on $\Omega_\hH$ gives the probability with which specific
configurations are realized. We now have the following stability
theorem showing that any quantity defined almost surely with respect to
$\PP$ is stable with respect to compact perturbations of the
Hamiltonian $\hH$.

\begin{theo}
\label{theo-stab}
Let $\hat{V}=\hat{V}^*$ be a compact operator. If $\hH$ is
homogeneous, so is $\hHV$. Then the symmetric difference of the
compact hulls $\Omega_\hH\triangle\Omega_\hHV$ is at most countable.
Moreover, any $T$-invariant measure on $\Omega_\hH$ or $\Omega_\hHV$
has its support in $\Omega_\hH\cap\Omega_\hHV$. Hence an invariant
measure on $\Omega_\hH$ completely determines an
invariant measure on $\Omega_\hHV$. 
\end{theo}

The proof is given in the appendix. In the sequel, we will drop the
index $\hH$ in $\Omega_\hH$ whenever there is no ambiguity.

\subsection{The non-commutative Brillouin zone}
\label{sec-Brillouin}

In the last section, we constructed the hull of the Hamiltonian $\hH$. Let
us briefly review the construction of the corresponding crossed
product \CS $\Aa$ called the non-commutative Brillouin zone (NCBZ) of
$\hH$ \cite{Be85,Be91}. For quasicrystals, the algebra is in general given
by a \CS associated to a groupoid \cite{Be91,Kel,Con}. In that case the
formul{\ae} below have direct analogs except for Birkhoff's theorem
(equation (\ref{eq-tracepervol})) which is not yet proved in that
context as far as we know. All the analysis of this article should
transpose directly to that case.

\vspace{.1cm}

Let us first consider the topological vector space
$\Cc_{\kappa}(\Omega \times  \ZZ^{d})$ of continuous functions with
compact support on $\Omega \times   \ZZ^{d}$.  It is endowed with the
following structure of a $^{\ast}$-algebra by

\begin{equation}
\label{eq-staralg}
AB ({\omega}, n) \; =\; \sum_{l\in\ZZ^{d}}
		A(\omega, l) B(T^{-l} \omega, n-l) 
			e^{  \frac{\imath q }{2\hbar }
			      \Bb.n \wedge l} 
				\mbox{ , }
\qquad
A^{\ast}(\omega, n) \; = \; \overline{A(T^{-n}\omega, -n)}
		\mbox{ , }	
\end{equation}

\noindent where $A,B\in \Cc_{\kappa}(\Omega \times  \ZZ^{d})$,
$\omega \in \Omega$, $n \in \ZZ^{d}$, finally  the antisymmetric real
tensor $\Bb=(B_{i,j})$ is a uniform magnetic field and $\Bb.n \wedge
l=\sum_{i,j}B_{i,j}n_il_j$. For $\omega \in \Omega$, this
$^{\ast}$-algebra is represented on $\HH=\ell^2(\ZZ^d)$  by 

\begin{equation}
\label{eq-repreg}
\pi_{\omega}(A)\psi (n)=\sum_{l\in\ZZ^{d}}
	A(T^{-n} \omega, l - n)
		e^{ \frac{\imath q }{2\hbar} 
		\Bb.l \wedge n } \psi (l) 
\mbox{ , }
\hspace{1cm}
\psi \in \ell^{2}(\ZZ^{d})
\mbox{ , }
\end{equation}

\noindent namely, $\pi_{\omega}$ is linear, $\pi_{\omega}(AB) =
\pi_{\omega}(A)\pi_{\omega}(B)$ and
$\pi_{\omega}(A)^{\ast}=\pi_{\omega} (A^{\ast})$. In addition,
$\pi_{\omega}(A)$ is a bounded operator.  Let a projective unitary
representations $(U(a))_{a\in\ZZ^d}$ on $\Hh$ be given by the magnetic
translations:

$$
U(a)\psi(n)\;=\;
e^{ \frac{\imath q }{2\hbar} \Bb.a \wedge n } \psi (n-a) 
\mbox{ . }
$$

\noindent Then the representations are related by the covariance condition 
$$
U({a})\pi_{\omega}(A)U({a})^{-1} =\pi_{T^{{a}}\omega}(A)\mbox{ , }
\qquad a\in\ZZ^d				\mbox{ . }
$$
\noindent Now $\parallel\! A\!\parallel =\sup_{\omega \in \Omega}
\parallel\! \pi_{\omega}(A)\!\parallel $ defines a $C^{\ast}$-norm.
This allows to define $\Aa=C^{\ast}(\Omega \times \ZZ^{d},\Bb )$
as the completion of $\Cc_{\kappa}(\Omega \times \ZZ^{d})$  under this
norm. Clearly, the  representations $\pi_\omega$ can be continuously
extended to this  $C^\ast$-algebra. This family of representations is
strongly continuous in $\omega$ for any fixed $A\in\Aa$. Finally,
there exists an element $H\in\Aa$ such that   $\pi_{\omega_0}(H)=\hH$
where $\omega_0$ is the point $\hH$ of $\Omega$ \cite{Be91}.

\vspace{.1cm}

Given an invariant and ergodic probability measure $\PP$ on
$\Omega$, a trace $\TV$ on all $\Aa$ is defined by

\begin{equation}
\label{eq-tracepervol}
\TV(A)\;=\;\int_{\Omega}d\PP(\omega)\;\langle
0|\pi_{\omega}(A)|0\rangle\;=\;
\lim_{l\rightarrow\infty}\frac{1}{|\Lambda_l|}\sum_{n\in\Lambda_l}
\langle n|\pi_{\omega'}(A)|n\rangle 
\mbox{ , }
\end{equation}

\noindent where $|n\rangle $ is the state completely localized at
$n\in\ZZ^d$.  The $\Lambda_l$'s are an increasing sequence of
rectangles centered at the origin. The equality holds for almost all
$\omega'$ by Birkhoff's ergodic theorem. This shows that $\TV$ is the
trace per unit volume. Note that a compact perturbation of the
Hamiltonian changes the \CS $\Aa$, but not the trace per unit volume of
observables.

$\TV$ gives rise to the GNS Hilbert space
$L^2(\Aa,\TV)$ and GNS representation $\pi_{\mbox{\tiny GNS}}$.  We
denote by $L^{\infty}(\Aa,\TV)$ the von Neumann algebra
$\pi_{\mbox{\tiny GNS}}(\Aa)''$ where $''$ is the bicommutant. By a
theorem of Connes \cite{BES}, $L^{\infty}(\Aa,\TV)$ is canonically
isomorphic to the von Neumann algebra of $\PP$-essentially bounded,
weakly measurable and covariant families $A_\omega$ of operators on
$\HH=\ell^2(\ZZ^2)$ endowed with the  norm

$$
\parallel\! A\!\parallel_{L^{\infty}}\;=\;\PP\!-\!\essinf_{\omega\in\Omega}
\parallel\! A_\omega\!\parallel_{\Bb(\HH)}
\mbox{ . }
$$

\noindent Consequently, the family of representations $\pi_\omega$
extends to a family of weakly measurable representations of
$L^{\infty}(\Aa,\TV)$. Moreover, the trace $\TV$ extends to 
$L^{\infty}(\Aa,\TV)$.

To define a differential structure on $\Aa$, consider the
family of $\ast$-automorphisms $\rho_{k_j}$ of $\Aa$ given by

$$
(\rho_{k_j} A)(\omega,{n})\;=\;
e^{\imath k_jn_j}A(\omega,{n})
\mbox{ , }
\qquad A\in\Aa\mbox{ . }
$$

\noindent Then the $d$ generators of $\rho_{k_j}$, denoted by
$\partial_j,\;\;j=1\ldots d$, are $\ast$-derivations. We use the
notation $\vec{\nabla}=(\partial_1\ldots \partial_d)$. 
If $\XV=(X_1,\ldots,X_d)$ is the position operator in $\HH=\ell^2(\ZZ^d)$, 

$$
(X_j\phi)(n)\;=\;n_j\phi(n)\mbox{ , }
\qquad \phi\in\ell^2(\ZZ^d)\mbox{ , }
n=(n_1,\ldots,n_d)\in\ZZ^d\mbox{ . }
$$

\noindent one can check that

$$
\pi_{\omega}(\rho_{k_j} (A))\;=\;
e^{\imath k_j {X_j}}(\pi_{\omega}(A))e^{-\imath k_j {X_j}}\mbox{ , }
$$

\noindent and
$\pi_{\omega}(\vec{\nabla}A)=\imath[\XV,\pi_{\omega}(A)]$. The
differential elements of $\Aa$ are

$$
\Cc^k(\Aa)\;=\;\{A\in\Aa\;|\;\partial_{j_1}\cdots\partial_{j_l}A\in\Aa,\;
j_1,\ldots, j_l\in 1,\ldots, d,\;l\leq k\}
\mbox{ . }
$$

\subsection{Local exponents of Borel measures}
\label{sec-expoess}

\begin{defini}
\label{def-diffexpointro}
Let $f$ and $g$ be Lebesgue-measurable non-negative functions on the
intervals $(0,b]$ and $[b,\infty)$ respectively, $b>0$. The behaviors
$f(x) \stackrel{{\textstyle \sim}}{{\scriptscriptstyle x\downarrow 0}}
x^{\au}$ and $g(x)  \stackrel{{\textstyle \sim}}{{\scriptscriptstyle
x\uparrow \infty}} x^{\eta}$ are defined by

\begin{equation}
\label{def-comp}
\au\;=\;\sup\{\gamma \in \RR \;|\;\exists\;\;a\in\RR:\; 0<a\leq b;
\; \int_0^a \frac{dx}{x}\;
\frac{f(x)}{x^{\gamma}}
\;<\; \infty \} \mbox{ , }
\end{equation}

\begin{equation}
\label{def-infty}
\eta=\inf\{\gamma \in \RR \;\mid\;\exists\;\;a\in\RR: \;b\leq
a<\infty;\;\;
 \int_a^{ \infty}
\frac{dx}{x}\;\frac{g(x)}{x^{\gamma}}
\;<\; \infty \} \mbox{ , }
\end{equation}
\end{defini}

\noindent Let $\Mm$ be the space of Borel probability measures on $\RR$
with the vague topology.

\begin{defini}
\label{def-spec} Let $\nu$ be in $\Mm$ and $E\in\RR$.   The local
exponent $\alpha_\nu(E)$ of $\nu$ at $E$ is defined by 
$$
\int^{E+\epsilon}_{E-\epsilon}d\nu(E')
\; \stackrel{{\textstyle \sim}}{{\scriptscriptstyle 
\epsilon\downarrow 0}} 
\;\epsilon^{\alpha_\nu(E)}
\mbox{ . }
$$
\end{defini}

\noindent Remark that the local exponents are always bigger than or equal
to $0$. Note further that Fubini's theorem immediately implies that 

\begin{equation}
\label{eq-expoca}
\alpha_{\nu}(E) 
\; = \; \sup\{\gamma\in\RR\;|
\;\;\int d\nu(E')\frac{1}{|E-E'|^{\gamma}}
\;<\;\infty\}\mbox{ . }
\end{equation}

\begin{theo}
\label{theo-expoppac} Let $\nu,\mu\in\Mm$.

\vspace{.1cm}

\noindent {\bf i)} For $\nu$-almost all $E$,
$0\leq\alpha_{\nu}(E)\leq 1$.

\vspace{.1cm}

\noindent {\bf ii)} If $\mu$ dominates $\nu$, then 
$\alpha_\mu(E)\leq\alpha_\nu(E)$ $\mu$-almost surely and
$\alpha_\nu(E)=\alpha_\mu(E)$ $\nu$-almost surely.

\vspace{.1cm}

\noindent {\bf iii)} If $\nu$ is pure-point, then 
$\nu$-almost surely $\alpha_\nu(E)=0$ .

\vspace{.1cm}

\noindent {\bf iv)} If $\nu$ is absolutely continuous, then 
$\alpha_{\nu}(E)=1$ ${\nu}$-almost surely.

\vspace{.1cm}

\noindent {\bf v)}
$(\nu,E)\in\Mm\times\RR\mapsto\alpha_{\nu}(E)$ is a Borel function.

\end{theo}

This result shows that $\nu$-almost surely in $E$, the local
exponent $\alpha_\nu(E)$ only depends upon the measure class of $\nu$.
Note that item {\bf ii)} does not follow directly from the
Radon-Nykodim theorem. For example, let $f(E)dE$, $\in L^1(\RR)$,  be
an absolutely continuous measure. The function $f$ may have
singularities where the exponent is smaller than $1$. However,
according to Theorem \ref{theo-expoppac}, these points only have
Lebesgue measure $0$. Note also that $\alpha_\nu(E)$ may be bigger
than $1$, but only on a set of zero $\nu$-measure. In practice, one
can use the following characterization which is an extension of the
Charles de la Vall\'ee Poussin theorem \cite{RS}. The proof can be
found in \cite{ZMB}, see also \cite{RJLS2}.

\begin{theo} 
\label{theo-stieltjes} {\rm \cite{ZMB}}
Let $\nu\in\Mm$ and $G_{\nu}(z)=\int_{\RR}\frac{d\nu(E')}{z-E'}$,
$\Im m(z)>0$, its Green's function.  The exponent $\beta(E)$ is
introduced by
$$
\Im m (G_{\nu}(E-\imath
\epsilon)) \;
\stackrel{{\textstyle \sim}}{{\scriptscriptstyle \epsilon\downarrow 0}} 
\;\epsilon^{\beta(E)}\mbox{ . }
$$

\noindent Then $\beta(E)=\au_{\nu}(E)-1$  whenever  $\au_{\nu}(E)\in
[0,2]$.
\end{theo}

The previous results allow to associate to a given measure
class $[\nu]$ a function $\alpha_\nu\in L^\infty(\RR,d\nu)$ taking
values in the interval $[0,1]$ $\nu$-almost surely. Of particular
interest are the biggest and smallest typical exponent in a given Borel
set (see \cite{You}). 

\begin{defini}
\label{def-specdim} Let $\nu\in\Mm$ and $\Delta\subset\RR$ be a Borel
set. The upper and lower essential exponents are defined by

$$
\alpha^+_{\nu}(\Delta)\;=\;
\nu\!-\!\esssup_{E\in\Delta}\alpha_\nu(E)
\mbox{ , }
\qquad
\alpha^-_{\nu}(\Delta)\;=\;
\nu\!-\!\essinf_{E\in\Delta}\alpha_\nu(E)
\mbox{ . }
$$

\end{defini}

\begin{coro}
\label{coro-essexpo}
Let $\mu,\nu$ be two Borel measures on $\RR$ and $\Delta\subset\RR$ a
Borel set. If $\mu$ dominates $\nu$, then $\alpha^+_{\nu}(\Delta)
\leq\alpha^+_{\mu}(\Delta)$ and $\alpha^-_{\mu}(\Delta)
\leq\alpha^-_{\nu}(\Delta)$. If $\mu$ and $\nu$ are in the same
measure class, then $\alpha^+_{\mu}(\Delta) =\alpha^+_{\nu}(\Delta)$
and $\alpha^-_{\mu}(\Delta) =\alpha^-_{\nu}(\Delta)$. 
\end{coro}

\begin{proposi}
\label{prop-essexpoborel}
Let $\Delta\subset\RR$ be a Borel set, then
$\nu\in\Mm\mapsto\alpha^+_\nu(\Delta)$ and $\nu\in\Mm
\mapsto\alpha^-_\nu(\Delta)$ are Borel functions.
\end{proposi}
 
The essential exponents are linked to the Hausdorff
dimensions $\mbox{dim}_{\mbox{\rm\tiny H}}$ (see for example
\cite{Fal}) associated to the measure $\nu$ and its support by the
following theorem of Rodgers and Taylor \cite{Rod} (our formulation
only slightly varies from theirs).

\vspace{.2cm}

\noindent {\bf Theorem} {\rm \cite{Rod}}
{\it Let $\nu\in\Mm$ and $\Delta\subset\RR$ Borel. If $S_0$ is the Borel
set defined by

$$
S_0\;=\;\{E\in\Delta\;|\; \alpha_\nu(E)\leq \alpha^+_\nu(\Delta)
\}
\mbox{ , }
$$

\noindent then $\nu(S_0)=\nu(\Delta)$ and $\mbox{\rm
dim}_{\mbox{\rm\tiny H}}(S_0)=\alpha^+_\nu(\Delta)$. There is no Borel
set $S\subset\Delta$ with $\mbox{\rm dim}_{\mbox{\rm\tiny
H}}(S)<\alpha^+_\nu(\Delta)$ satisfying $\nu(S)=\nu(\Delta)$.
Moreover, if a Borel set $S\subset\Delta$ satisfies $\mbox{\rm
dim}_{\mbox{\rm\tiny H}}(S)<\alpha^-_\nu(\Delta)$, then  $\nu(S)=0$.}

\vspace{.2cm}

The Hausdorff dimension of a measure is defined by \cite{You}

$$
\mbox{\rm dim}_{\mbox{\rm\tiny H}}(\nu|_\Delta)\;=\;
\inf_{S\subset\Delta}\;\{
\mbox{\rm dim}_{\mbox{\rm\tiny H}}(S)\;|\;\nu(S)=\nu(\Delta)\}
\mbox{ , }
$$

\noindent where the infimum is taken over Borel sets $S\subset\Delta$.
Rodgers' and Taylors theorem shows that  $\mbox{\rm
dim}_{\mbox{\rm\tiny H}}(\nu|_\Delta)= \alpha^+_\nu(\Delta)$.

\subsection{Local exponents of a self-adjoint operator}
\label{sec-expoop}

Let $H$ be a selfadjoint operator acting on the separable Hilbert
space $\HH$. The spectral theory associates to $H$ a
$\HH$-projection-valued Borel measure $\Pi$ on $\RR$ \cite{RS}.
Furthermore, for any $\phi\in\HH$, $\parallel\!\phi\!\parallel=1$, let
$\rho_\phi$ be the spectral measure of $H$ relative to $\phi$, namely
for $f\in C_0(\RR)$,

$$
\int d\rho_\phi(E)\;f(E)\;=\;\langle\phi|f(H)|\phi\rangle
\;=\;\int\langle\phi|\Pi(dE)|\phi\rangle\;f(E)
\mbox{ . }
$$

\noindent In physics literature, $\rho_\phi$ is called the {\sl local
density of states} (LDOS).

\begin{defini}
\label{def-spec2} 
Let $E\in\RR$ and $\Delta$ be a Borel subset of $\RR$. The spectral
exponent and essential spectral exponents of $\Pi$ {\rm (}or $H${\rm
)} are defined by

$$
\alpha_{\Pi}(E)\;=\;\inf_{\phi\in\HH}\alpha_{\rho_{\phi}}(E)\mbox{ , }
\qquad
\alpha^+_{\Pi}(\Delta)\;=\;
\sup_{\phi\in\HH}\alpha^+_{\rho_\phi}(\Delta)
\mbox{ , }
\qquad
\alpha^-_{\Pi}(\Delta)\;=\;
\inf_{\phi\in\HH}\alpha^-_{\rho_\phi}(\Delta)
\mbox{ . }
$$

\end{defini}

\begin{theo}
\label{theo-Piexpogene}
There exists $\psi\in\HH$ with

$$
\alpha^+_{\Pi}(\Delta)\;=\;
\alpha^+_{\rho_\psi}(\Delta)
\mbox{ , }
\qquad
\alpha^-_{\Pi}(\Delta)\;=\;
\alpha^-_{\rho_\psi}(\Delta)\mbox{ . }
$$

\end{theo}

This result shows that there are {\sl typical} states in
$\Hh$ giving the generic properties of the spectrum.

\subsection{Density of states and local density of states}
\label{sec-DOSLDOS}

Let $H\in\Aa$ be the Hamiltonian. The spectral projection of
$\pi_\omega(H)$ is denoted by $\Pi_\omega$.

\begin{theo}
\label{theo-dimLDS}
For $E\in\RR$ and a Borel subset $\Delta\subset\RR$, the exponents
$\alpha_{\Pi_{\omega}}(E)$, $\alpha_{\Pi_{\omega}}^+(\Delta)$ and
$\alpha_{\Pi_{\omega}}^-(\Delta)$ are  $\PP$-almost surely independent
of $\omega$. The common values are denoted by $\alpha_{\mbox{\rm\tiny
LDOS}}(E)$, $\alpha^+_{\mbox{\rm\rm\tiny LDOS}}(\Delta)$ and
$\alpha^-_{\mbox{\rm\rm\tiny LDOS}}(\Delta)$ respectively.
\end{theo}

\noindent A related result was proved by Last \cite{Las}. Theorem
\ref{theo-stab} implies that all these exponents are stable with
respect to compact perturbation of the Hamiltonian.

\begin{coro}
If $\pi_{\omega}(H)$ has pure-point spectrum in $\Delta$, $\PP$-almost
surely, then  $\alpha^{\pm}_{\mbox{\rm\tiny LDOS}}(\Delta)=0$.  If
$\pi_{\omega}(H)$ has absolutely continuous spectrum in $\Delta$ for
$\PP$-almost all $\omega\in\Omega$, then
$\alpha^{\pm}_{\mbox{\rm\tiny LDOS}}(\Delta)=1$. If
$0<\alpha^-_{\mbox{\rm\tiny LDOS}}(\Delta)\leq
\alpha^+_{\mbox{\rm\tiny LDOS}}(\Delta)<1$, then the spectrum is
singular continuous  in $\Delta$ for $\PP$-almost all
$\omega\in\Omega$.
\end{coro}

Another important spectral measure associated to $H$ is the
density of states (DOS) defined by

\begin{equation}
\label{eq-DOSdef}
\int d\Nn(E)\;f(E)\;=\;\int d\PP(\omega)\,\langle
0|\pi_\omega(f(H))|0\rangle
\mbox{ , }
\qquad
f\in C_0(\RR)\mbox{ . }
\end{equation}

\noindent We denote $\alpha_{\mbox{\rm\rm\tiny
DOS}}(E)=\alpha_{\Nn}(E)$ and for any Borel subset $\Delta$ of $\RR$,
we set:

$$
\alpha^\pm_{\mbox{\rm\rm\tiny DOS}}(\Delta)\;=\;
\alpha_{\Nn}^\pm(\Delta)\mbox{ , }
$$

\begin{theo}
\label{theo-LDSDOS}
For $E\in\RR$ and a Borel subset $\Delta\subset\RR$, 
$$
\alpha_{\mbox{\rm\tiny LDOS}}(E)\;\leq\;
\alpha_{\mbox{\rm\rm\tiny DOS}}(E)\mbox{ , }
\qquad
\alpha^\pm_{\mbox{\rm\rm\tiny LDOS}}(\Delta)\;\leq\;
\alpha^\pm_{\mbox{\rm\rm\tiny DOS}}(\Delta)\mbox{ , }
$$
\end{theo}

Note that this implies the Hausdorff dimension of the DOS is
bigger than or equal to the Hausdorff dimension of the LDOS.

\subsection{Diffusion exponents and localization}
\label{sec-diffuexpo}

This section is devoted to dynamical quantum diffusion and quantum
localization. The diffusion exponent allows to measure the importance of
quantum interference effects due to the frozen (disorder or
quasiperiodic) potential in the one-particle Hamiltonian. Collisions
with time-dependent disorder such as collisions with phonons and its
effects on diffusion are not considered here; in RTA 
these effects are treated by the phenomenological
constant $\trel$ in Kubo's formula. Diffusion is supposed to be
isotropic here, but this is only done for sake of notational
simplicity.

For a given Borel set $\Delta\subset \RR$, the mean
square displacement operator is

\begin{equation}
\label{def-Lsqdisproj}
\delta X^2_{\omega,\Delta}(T) \;{=}\; \int_0^T\frac{dt}{T}\;
\Pi_{\omega}(\Delta)(\vec{X}_{\omega}(t)-\vec{X})^2\Pi_{\omega}(\Delta) 
\mbox{ , }
\end{equation}

\noindent where $\vec{X}_{\omega}(t)=
e^{\imath t\pi_\omega(H)}\vec{X}e^{-\imath t\pi_\omega(H)}$. 

\begin{defini}
\label{def-exdiff}
The diffusion exponent $\sigma_{\mbox{\rm\tiny diff}}(\Delta)$ is
defined by

\begin{equation}
\label{def-diffexpo1}
\int_{\Omega}d\PP(\omega)\;\langle 0|\delta
X^2_{\omega,\Delta}(T)|0\rangle \;
\stackrel{{\textstyle \sim}}{{\scriptscriptstyle T\uparrow \infty}} 
\;T^{2\sigma_{\mbox{\rm\tiny diff}}(\Delta)}
\mbox{ . }
\end{equation}
\end{defini}

\begin{proposi}
\label{prop-exdiff3}
Let $\Pi(\Delta)=\chi_\Delta(H) \in
L^\infty(\Aa,\TV)$ where $\chi_\Delta$ is the characteristic function
on the Borel set $\Delta$ and suppose that $H\in\Cc^1(\Aa)$. Then 

\begin{equation}
\label{def-exdiff3}
\int_0^T\frac{dt}{T}\;
 \TV (|\vec{\nabla}( e^{-\imath Ht})|^2\Pi({\Delta}))
  =
   \int_{\Omega}d\PP(\omega)\;\langle 0|\delta
    X^2_{\omega,\Delta}(T)|0\rangle \;
\;\stackrel{{\textstyle \sim}}{{\scriptscriptstyle T\uparrow \infty}} 
 \;T^{2\sigma_{\mbox{\rm\tiny diff}}(\Delta)}
\mbox{ . }
\end{equation}
\end{proposi}

\begin{theo}
\label{theo-diffgener} Let $H\in\Cc^1(\Aa)$. Then:

\vspace{.1cm}

\noindent {\bf i)} $0\leq \sigma_{\mbox{\rm\tiny diff}}(\Delta)\leq 1$.

\vspace{.1cm}

\noindent {\bf ii)}  Let $\hat{V}$ be a compact operator on
$\Hh=\ell^2(\ZZ^d)$ such that $[\XV,\hat{V}]$ is bounded. Let the
invariant ergodic measure on $\Omega_\hH$ determine that on
$\Omega_\hHV$ as in Theorem {\rm \ref{theo-stab}}, then the diffusion
exponents $\sigma_{\mbox{\rm\tiny diff}}(\Delta)$ of $\hH$ and
$\hH+\hat{V}$ are equal. 

\vspace{.1cm}

\noindent {\bf iii)}  Guarneri's bound: for any open interval
$\Delta\subset \RR$:

\begin{equation}
\label{eq-Guarn}
{\alpha}^+_{\mbox{\rm\tiny LDOS}}(\Delta)\;\leq\;
d\cdot\sigma_{\mbox{\rm\tiny diff}}(\Delta)
\mbox{ , }
\end{equation}

\noindent whenever $H\in\Cc^k(\Aa)$ for some $k>d/2$.
\end{theo}

An inequality between spectral and diffusion exponents was
first proved by Guarneri \cite{Gua}. A further contribution is due to
Combes \cite{Com}. Last improved the proof in order to show that it is
the most continuous part of the spectrum which gives the lower bound
of the diffusion exponent \cite{Las}, see also \cite{BCM}.  The bound
(\ref{eq-Guarn}) links exponents associated to the covariant family of
Hamiltonians irrespective of the choice of a specific vector in
Hilbert space.

\vspace{.2cm}

Let us conclude this section with a discussion of localization. The
following localization criterion for a Borel subset $\Delta\subset\RR$ 
was introduced in
\cite{Be94} motivated by the study of the quantum Hall effect
\cite{BES}: 
 
\begin{equation}
\label{def-lo}
l^2 (\Delta)\;=\;
\limsup_{T\rightarrow \infty}
\int_0^T\frac{dt}{T}\;\TV(|\vec{\nabla}e^{-\imath Ht}|^2\;\Pi(\Delta))
\;<\;\infty 
\mbox{ . }
\end{equation}

Note that it is strictly stronger than $\sigma_{\mbox{\tiny
diff}}(\Delta)=0$ because no logarithmic divergencies are allowed.
Actually, (\ref{def-lo}) coincides with the localization criterion used
by physicists: in physics literature, averages of products of Green
functions are used; this leads to the current-current correlation
measure $m$ below (Theorem \ref{theo-locexpr}). In the Anderson
model and a wide class of other models, the condition (\ref{def-lo})
has been shown to hold for the spectral subsets generally considered to
be localized \cite{BES}.

\begin{theo}
\label{theo-locpoint} Suppose that the localization condition {\rm
(\ref{def-lo})} is satisfied for a Borel set $\Delta\subset\RR$. Then the
following holds:

\vspace{.1cm}

\noindent {\bf i)}  {\rm \cite{Be94,BES}} 
$\sigma_{\mbox{\tiny diff}}(\Delta)=0$ and $\pi_{\omega}(H)$ has
pure-point spectrum in $\Delta$ for  $\PP$-almost every
$\omega\in\Omega$.

\vspace{.1cm}

\noindent {\bf ii)} Let $\hat{V}$ be a compact self-adjoint operator
such that $[\XV,\hat{V}]$ is bounded.  Let the invariant ergodic
measure on $\Omega_\hH$ determine that on $\Omega_\hHV$ as in Theorem
{\rm \ref{theo-stab}}, then the localization condition {\rm
(\ref{def-lo})}  is simultaneously satisfied for $\hH$ and
$\hH+\hat{V}$. 

\vspace{.1cm}

\noindent {\bf iii)}  {\rm \cite{Be94,BES}} There is a
${\cal N}$-measurable function $l$ on $\Delta$ such that for every
Borel subset $\Delta'$ of $\Delta$:

\begin{equation}
\label{eq-loclen}
l^2 (\Delta')=\int_{\Delta'}d{\cal N}(E)\; l(E)^2
\mbox{ . }
\end{equation}

\end{theo} 

Let us notice that the criterion (\ref{def-lo}) can be
weakened in the following way: let $g$ be any increasing function on
$\RR^+$ such that $\lim_{x\rightarrow\infty}g(x)=\infty$ and consider

$$
l_g(\Delta)\;=\;
\limsup_{T\rightarrow \infty}
\int_0^T\frac{dt}{T}\;
\int_\Omega d\PP(\omega)\;\langle 0|\Pi_\omega(\Delta)
g(|\XV_\omega(t)-\XV|)\Pi_\omega(\Delta)|0\rangle
\mbox{ . }
$$

\noindent Then $l_g(\Delta)<\infty$ suffices to get pure-point
spectrum in $\Delta$, $\PP$-almost surely and to insure that this
property is stable by compact perturbations of the Hamiltonian.

\subsection{Current-current correlation function}
\label{sec-curcur}

In this section we give some useful formul{\ae} for the calculation of
the diffusion exponent. As illustrative application, the diffusion
exponent of Wegner's $n$-orbital model is calculated in Chapter
\ref{chap-Andfree}. The current operator is defined (if
$H\in\Cc^1(\Aa)$) by

$$
\vec{J}\;=\;\vec{\nabla}(H)\mbox{ . }
$$

\noindent The current-current correlation functions are the Borel measures
$m_{i,j}$ on $\RR^2$ given by \cite{KP}

\begin{equation}
\label{eq-defmes}
\int_{\RR^2} dm_{i,j}(E,E')\;f(E)g(E')\;=\;\TV(\partial_i(H)f(H)
\partial_j(H)g(H))\mbox{ , }
\end{equation}

\noindent where $f,g\in C_0(\RR)$. The right hand side
defines a positive and continuous bilinear form on $C_0(\RR)\times
C_0(\RR)\times M_d(\CC)$. The Riesz-Markov theorem \cite{RS} then assures the
existence of the Radon measures $m_{i,j}$ on $\RR^2$ with finite mass.
The cyclicity of the trace induces the following symmetry of $m_{j,j}$
with respect to the diagonal $E=E'$:

\begin{equation}
\label{eq-sym}
\int_{\RR^2} dm_{j,j}(E,E')\;f(E,E')\;=\;
\int_{E\geq E'} dm_{j,j}(E,E')\;(f(E,E')+f(E',E))\mbox{ . }
\end{equation}

\noindent The isotropic part $m$ is the measure $m=\sum_{j=1}^dm_{j,j}/d$. It is
called the current-current correlation measure or also the conductivity measure.
It allows to calculate the diffusion exponent.

\begin{theo}
\label{theo-exdiff} Given a Borel set $\Delta\subset\RR$ and
$\epsilon>0$, let {\rm diag}$(\Delta,\epsilon)$ be the set of points
in $\Delta\times\RR$ within distance $\epsilon$ from the diagonal in
$\RR^2$, then

$$
\int_{\mbox{\tiny diag}(\Delta,\epsilon)}dm(E,E')
\;
\stackrel{{\textstyle \sim}}{{\scriptscriptstyle \epsilon\downarrow 0}} 
\;
\epsilon^{2(1-\sigma_{\mbox{\tiny\rm diff}}(\Delta))}\mbox{ . }
$$
\end{theo}

\noindent The Stieltjes transform of $m$ is given by

\begin{equation}
\label{eq-condcalc}
S_m(z_1,z_2) \; = \; 
\frac{1}{(2\pi\imath)^2}\;\int_{\RR^2}dm(E,E')\;
\frac{1}{(E-z_1)(E'-z_2)}
\mbox{ . }
\end{equation}

\noindent If $H=H_0+V$ with a translation invariant kinetic part
$H_0$ and a potential $V$ satisfying $\vec{\nabla}(V)=0$, then
$S_m$ can be calculated by means of the 2-point Green's function:

\begin{equation}
\label{eq-2poin}
S_m(z_1,z_2) \:= \; \frac{1}{d}\;
\frac{1}{(2\pi\imath)^2}\;\sum_{r,s,t\in\ZZ^d}
\langle 0|\vec{\nabla}(H_0)|r\rangle \cdot\langle s|\vec{\nabla}(H_0)|t\rangle \;
G^2(z_1,z_2,r,s,t,0)\mbox{ , }
\end{equation}

\noindent where

\begin{equation}
\label{eq-twopart}
G^2(z_1,z_2,r,s,s',r')\;=\;
\int_{\Omega} d\PP(\omega)\;\langle r|\frac{1}{z_1-\pi_{\omega}(H)}|s\rangle 
\langle s'|\frac{1}{z_2-\pi_{\omega}(H)}|r'\rangle 
\mbox{ . }
\end{equation}

\begin{theo}
\label{theo-exdiff2}
The diffusion exponent is given by

$$
\Re e \left(\int_{\RR}da\;S_m(a+\imath {\epsilon},
a-\imath {\epsilon})\right)\;
\stackrel{{\textstyle \sim}}{{\scriptscriptstyle \epsilon\downarrow
0}}
\;\epsilon^{1-2\sigma_{\mbox{\rm\tiny diff}}(\RR)}
\mbox{ . }
$$

\end{theo}
 
\noindent The localization criterion can also be expressed by means of the
conductivity measure \cite{BES}

\begin{theo}
\label{theo-locexpr}
{\rm \cite{BES}} The localization condition
{\rm (\ref{def-lo})} is equivalent to

$$
\int_{\Delta\times\RR}dm(E,E')\;\frac{1}{|E-E'|^2}\;<\;\infty
\mbox{ . }
$$

\end{theo}

Let us define the Liouville operator $\Ll_H$ acting on $A\in\Aa$ by
$\Ll_H(A)=\imath [H,A]/\hbar$. Then the spectral measure $\rho_{\vec{J}}$ of
$\imath\Ll_H$ associated to the current operator $\vec{J}$ is defined
by (for $f\in C_0(\RR)$)

\begin{equation}
\label{def-Liou}
\int d\rho_{\vec{J}}(\epsilon) 
\;f(\epsilon)\;=\;\frac{1}{d}\;
\TV(f(\imath\Ll_H)(\vec{J})\cdot
\vec{J})
\mbox{ , }
\end{equation}

\begin{theo}
\label{theo-Liou}
The spectral exponent $\alpha_{\rho_{\vec{J}}}(0)$ is given by

$$
\alpha_{\rho_{\vec{J}}}(0)
\;=\;2(1-
\sigma_{\mbox{\rm\tiny diff}}(\RR))
\mbox{ . }
$$ 

\end{theo}

\subsection{Anomalous Drude Formula}
\label{sec-Kubo}

In \cite{BES}, we showed that the zero frequency,
isotropic direct conductivity at inverse temperature
$\beta$, chemical potential $\mu$ and relaxation time 
$\tau_{\mbox{\rm\tiny rel}}$ is given by

\begin{equation}
\label{eq-kubo3}
\sigma_{\beta,\mu}\;=\; \frac{2q^2}{\tau_{\mbox{\rm\tiny rel}}
 \hbar^2}\int_{E\geq E'}dm(E,E')\;
\frac{f_{\beta,\mu}(E')-f_{\beta,\mu}(E)}{E-E'}\;
\frac{1}{\frac{1}{\tau^2_{\mbox{\rm\tiny rel}}}+
\left(\frac{E-E'}{\hbar}\right)^2}
\mbox{ . }
\end{equation}

\noindent Here $f_{\beta ,\mu}(E)$ is
the Fermi-Dirac function $(1+e^{\beta(E-\mu)})^{-1}$, $q$ is the
particle charge and $\hbar$ is Planck's constant. More details on the
derivation of (\ref{eq-kubo3}) will be given in a forthcoming work.

\begin{theo}
\label{theo-Drudeano} If $\beta<\infty$, the direct conductivity
given in {\rm (\ref{eq-kubo3})} satisfies

\begin{equation}
\label{eq-comp1}
\sigma_{\beta,\mu}\;
\stackrel{{\textstyle \sim}}{{\scriptscriptstyle \tau_{\mbox{\rm\tiny rel}}\uparrow
\infty}} \;\tau_{\mbox{\rm\tiny rel}}^{-1+2\sigma_{\mbox{\rm\tiny
diff}}(\RR)}
\mbox{ . }
\end{equation}

\end{theo}

If $\trel\sim\beta^\alpha$ with $\alpha\approx 1-5$ as indicated in the
introduction, a more detailed analysis shows that only exponents
at the Fermi level $\mu$ intervene in the anomalous Drude formula.

\vspace{.5cm}

\section{Exponents: generalities}
\label{chap-gener}

\subsection{Local regularity behavior}

In this section we compare different exponents characterizing the
H\"older regularity behavior of positive functions. Although not all
of these exponents will be used in this article, we present them for
sake of completeness and later reference.

\begin{defini}
\label{def-diffexpo}
Let $f$  be a Lebesgue-measurable non-negative functions on the
interval  $(0,b]$, $b>0$. The exponents $\hat{\beta} $ and $\beta$ are
defined by

\begin{equation}
\label{def-limsupinf}
\hat{\beta}\;=\;\limsup_{x\rightarrow 0} \frac{\mbox{\rm Log }
f(x)}{\mbox{\rm Log }x}\mbox{ , }
\qquad
\beta\;=\;\liminf_{x\rightarrow 0} \frac{\mbox{\rm Log }
f(x)}{\mbox{\rm Log }x}
\mbox{ . }
\end{equation}

\end{defini}

\begin{rem} {\rm By convention a function vanishing in a neighborhood
of the origin will have exponents equal to infinity.}
\hfill $\diamond$
\end{rem}

\begin{proposi} 
\label{prop-diffexpo}
Let $f$ and $g$ be Lebesgue-measurable non-negative functions on the
interval  $(0,b]$, $b>0$.  Let $\alpha$, $\hat{\beta}$ and $\beta$ be
as in {\rm Definitions \ref{def-diffexpointro}} and {\rm
\ref{def-diffexpo}}.

\vspace{.1cm}

\noindent {\bf i)} 
If $f(x) \stackrel{{\textstyle \sim}}{{\scriptscriptstyle x\downarrow
0}}  x^{\au}$, then  $f(x)\mbox{\rm Log}(x) \stackrel{{\textstyle
\sim}}{{\scriptscriptstyle x\downarrow 0}}  x^{\au}$.  

\vspace{.1cm}

\noindent {\bf ii)} {\rm (Calculation with Laplace transform)}
If $f(x) \stackrel{{\textstyle \sim}}{{\scriptscriptstyle x\downarrow
0}}  x^{\au}$ and $\alpha>-1$, then

$$
\int^1_0 dt\;e^{-\delta t}f(t)\;
\stackrel{{\textstyle \sim}}{{\scriptscriptstyle \delta\uparrow \infty}} \;
\delta^{-\alpha-1}
\mbox{ . }
$$

\vspace{.1cm}

\noindent {\bf iii)} {\rm \cite{Hol}} The following equalities hold:

\begin{equation}
\label{def-liminf}
\beta\;=\;\sup \{\gamma\in \RR\;|\;
\limsup_{x\downarrow 0}\frac{f(x)}{x^\gamma}\;<\;\infty
\}\mbox{ , }
\qquad
\hat{\beta}\;=\;\inf \{\gamma\in \RR\;|\;
\liminf_{x\downarrow 0}\frac{x^\gamma}{f(x)}\;<\;\infty
\}\mbox{ . }
\end{equation}

\vspace{.1cm}

\noindent {\bf iv)}  $\beta\;\leq\;\au\;\leq\;\hat{\beta}$.

\vspace{.1cm}

\noindent {\bf v)} For $\au>0$ and  $f$  
non-decreasing {\rm (}respectively, for $\au\leq 0$ and  $f$  
non-increasing{\rm )}, $\beta=\au$.

\vspace{.1cm}

\noindent {\bf vi)} Suppose that both $f$ and $g$ are
non-increasing or non-decreasing with corresponding exponents
$\au_f$ and $\au_g$ 
as defined in {\rm (\ref{def-comp})}.
Then for $\delta>0$,

\begin{equation}
\label{eq-expon}
f(x)^{\delta}\; \stackrel{{\textstyle \sim}}{{\scriptscriptstyle 
x\downarrow 0}} 
\; x^{\delta\alpha_f}\mbox{ , }
\qquad
f(x)g(x)\; \stackrel{{\textstyle \sim}}{{\scriptscriptstyle 
x\downarrow 0}} 
\; x^{\alpha_f+\alpha_g}\mbox{ , }
\qquad
f(x)+g(x)\; \stackrel{{\textstyle \sim}}{{\scriptscriptstyle 
x\downarrow 0}} 
\; x^{\min\{\alpha_f,\alpha_g\}}\mbox{ . }
\end{equation}

\end{proposi}

\begin{rem} {\rm
These results transpose directly to the study of the
behavior of a function at infinity as given in Definition 
\ref{def-diffexpointro}. Note that in particular, if
$g(x) \stackrel{{\textstyle \sim}}{{\scriptscriptstyle x\uparrow
\infty}}x^{\alpha}$, $\alpha>-1$, 
and $I(\delta)=\int^\infty_1 dx \,e^{-\delta x}
g(x)$, then Proposition \ref{prop-diffexpo}ii) and iii) show that}

$$
\alpha\;=\;\inf
\{ \gamma\in\RR\;|\;\limsup_{\delta\downarrow
0}\delta^{\gamma+1}I(\delta)=0\}
\mbox{ . }
$$
\hfill $\diamond$
\end{rem}

\begin{rem} {\rm
The following example will show that there exist functions with
$\beta<\au<\hat{\beta}$ and for which the conclusions 
of Proposition \ref{prop-diffexpo}v)
do not hold. Let $t>s>1$, $u\in\RR$, and consider
\begin{equation}
\label{eq-example}
f(x)\;=\;
\left\{ \begin{array}{cc} n^u \;\;\;\;\;\;\; &
\mbox{for }\;\; x\in I_n=[\frac{1}{n^s},\frac{1}{n^s}+
\frac{1}{n^{s+t}}]
\mbox{ , } \\
0 \;\;\;\;\;\;\;\;\; & \mbox{otherwise . }
\;\;\;\;\;\;\;\;\;\;\;\;\;\;\;\;\;\;\;\;\;\;\;
 \end{array}
\right.
\end{equation}

\noindent Because of (\ref{def-liminf}) we have
$\hat{\beta}=\infty$ and $\beta=-\frac{u}{s}$. By explicit 
calculation one gets
\begin{equation}
\label{eq-excal}
\au\;=\; \frac{t-1}{s}-\frac{u}{s} \mbox{ . }
\end{equation}

\noindent As an example, take  $u=0$,
$s=2$ and  $t=5$, then $\beta=0$, $\au=2$ and $\hat{\beta}=\infty$. 
To consider the function $f^{\delta}$ is
equivalent to replacing $u$ by $u\delta$ and this leads, according to
(\ref{eq-excal}), to a exponent
different from $\delta\alpha$.}
\hfill $\diamond$
\end{rem}

\noindent {\bf Proof} of Proposition \ref{prop-diffexpo}.
{\bf i)} This follows from the fact that $\int^1_0dx\,
x^{-1+\epsilon}\mbox{Log}\,x\,<\infty$ for any $\epsilon>0$.

\noindent {\bf ii)} For $0>\gamma>-\alpha-1$, the identity

$$
\int^\infty_1\frac{d\delta}{\delta^{1+\gamma}}\;
\int^1_0 dt\;e^{-\delta t}f(t)\;=\;
\int^1_0 dt\;f(t)t^\gamma\int^\infty_t\frac{ds}{s^{1+\gamma}}\,e^{-s}
$$

\noindent allows to conclude.

\noindent {\bf iii)} is proved in \cite{Hol}.  

\noindent {\bf iv)} Let us only show $\beta\leq\alpha$. The other
inequality can be proved in a similar way. For  any given $\delta > 0$
there is a $\epsilon(\delta)\leq 1$ such that

$$
\inf_{x\leq\epsilon(\delta)} 
\frac{\mbox{Log }f(x)}{\mbox{Log }x}
\;=\;{\beta} - \delta
\mbox{ . }
$$

\noindent Then for $x\leq\epsilon(\delta)$, $f(x)\leq x^{{\beta} -
\delta}$ because $\mbox{Log}\;x<0$.  Let now $\gamma<{\beta}$ and
choose $\delta$ such that ${\beta} -\delta - \gamma>0$, then

$$
\int_0^{\epsilon(\delta)} \frac{dx}{x}\; 
\frac{f(x)}{x^{\gamma}}\;\leq\;
\int_0^{\epsilon(\delta)} \frac{dx}{x}\;
x^{{\beta} -\delta - \gamma}\;<\;\infty
\mbox{ , }
$$

\noindent which shows $\gamma\leq{\alpha}$.

\noindent {\bf v)} We only treat the case where $\au>0$ and $f$ is
non-decreasing. Take  $0<\gamma<\au$, then if $x\leq a/2$

$$
C(\gamma)\;\;=\;\;
\int_0^a \frac{dy}{y}\;\frac{f(y)}{y^{\gamma}}\;\;\geq\;\;
\int_x^{2x} \frac{dy}{y}\;\frac{f(y)}{y^{\gamma}}\;\;\geq\;\;
f(x)\int_x^{2x} \frac{dy}{y^{1+\gamma}}\;\;\geq\;\;
\frac{f(x)}{x^{\gamma}}\frac{1}{\gamma}(1-\frac{1}{ 2^{\gamma}})
\mbox{ , }
$$

\noindent and therefore equality (\ref{def-liminf}) implies that 
$\gamma\leq\beta$ and hence $\au\leq\beta$. Thanks to {\bf iv)}
this gives $\au=\beta$.

\noindent {\bf vi)} is a direct consequence of {\bf iii)} and {\bf v)}.
\hfill $\Box$

\vspace{.2cm}

\subsection{Local exponents and essential exponents}
\label{sec-locexpo}

We begin this section with the proof of Theorem \ref{theo-expoppac}.
Then follow some comments on Definitions \ref{def-spec} and
\ref{def-specdim} and Theorem \ref{theo-expoppac}. In the rest of the
section we prove the other results of Sections \ref{sec-expoess} and
\ref{sec-expoop} as well as some complementary results.

The following lemma is known as the Hardy-Littlewood maximal
inequality. We will need it in a slightly generalized form,
nevertheless, its proof can be directly transposed from \cite{Rud}, for
example.

\begin{lemma} 
\label{lem-maximal} 
Let $\mu,\nu$ be two probability measures on $\RR$ and $h\in
L^1(\RR,d\nu)$. The maximal function $M_{\mu,\nu,h}$ is defined by 

$$
M_{\mu,\nu,h}(E)\;=\;\sup_{\epsilon\in (0,1]}\;
\frac{1}{\mu([E-3\epsilon,E+3\epsilon])}\;
\int_{(E-\epsilon,E+\epsilon)}d\nu(E')\;h(E')
\mbox{ . }
$$ 

\noindent It is lower semicontinuous and satisfies for any positive
$\lambda$:

$$
\mu(\{E\in\RR\;|\;M_{\mu,\nu,h}(E)>\lambda\})\;\leq\;\frac{1}{\lambda}
{\parallel\! h \!\parallel_{L^1(\RR,d\nu)}}
\mbox{ . }
$$

\end{lemma}

\begin{lemma} 
\label{lem-ac} Let $\mu,\nu$ be two probability measures on $\RR$. 
Then $\alpha_\mu(E)\leq \alpha_\nu(E)$ $\mu$-almost surely.
\end{lemma}

\noindent {\bf Proof.}  If $M_{\mu,\nu,{\bf 1}}(E)<\infty$, then
${\nu((E-\epsilon,E+\epsilon))}<C{\mu([E-3\epsilon,E+3\epsilon])}$ for
all $\epsilon\in(0,1]$ and some constant $C>0$. Therefore
$\alpha_\mu(E)\leq \alpha_\nu(E)$. Thus $\alpha_\nu(E)< \alpha_\mu(E)$
implies $M_{\mu,\nu,{\bf 1}}(E)=\infty$. Hence by Lemma
\ref{lem-maximal}

$$
\mu(\{E\in\RR\;|\;\alpha_\nu(E)<\alpha_\mu(E) \}) \; \leq \;
\mu(\bigcap_{N\in\NN}\{E\in\RR\;|\;M_{\mu,\nu,{\bf 1}}(E)>N\})
\;\leq\;\lim_{N\rightarrow\infty}\frac{1}{N}\;=\;0
\mbox{ . }
$$
\hfill $\Box$

\vspace{.2cm}

\noindent {\bf Proof} of Theorem \ref{theo-expoppac}. {\bf i)} Clearly
the local exponents are all bigger than or equal to $0$. The exponents
of the Lebesgue measure are all equal to $1$. Applying Lemma
\ref{lem-ac} to the measure $\nu$ and the Lebesgue measure shows that
$\alpha_\nu(E)\leq 1$ for $\nu$-almost all $E\in\RR$.

{\bf ii)} Apply Lemma \ref{lem-ac} twice and use that
$\mu$-almost surely implies $\nu$-almost surely.

{\bf iii)} Since $\nu$ is pure-point, it is of the form
$\sum_{n\in\NN} c_n\delta(E-E_n)$, $c_n>0$. For each $E_n$,
$\nu([E_n-\epsilon, E_n+\epsilon])\geq c_n$ such that
$\alpha_{\nu}(E_n)=0$. Consequently $\alpha_{\nu}(E)$ is equal to $0$
for $\nu$-almost all $E$, notably the $E_n$'s.

{\bf iv)} If $\nu$ is absolutely continuous, it is dominated
by the Lebesgue measure. {\bf ii)} allows to conclude.

{\bf v)} We will prove a stronger result in Proposition
\ref{prop-Piexpogene}iii) below. \hfill $\Box$

\vspace{.2cm}

\begin{rem} {\rm 
Proposition \ref{prop-diffexpo} implies that the exponents
$\alpha_\nu(E)$ are the same as those often considered in literature
\cite{You,Cut,Pes,Ols,Las,RJLS2,BCM} because
$\nu([E-\epsilon,E+\epsilon])$ is a non-decreasing function of
$\epsilon$. }
\hfill $\diamond$
\end{rem}

\begin{rem} {\rm Theorem \ref{theo-expoppac}  does not exclude
singular continuous spectrum with exponents equal to $0$ or $1$.}
\hfill $\diamond$
\end{rem}

\begin{rem} {\rm An absolutely continuous measure can have exceptional
points where the exponent is not equal to $1$. For example, consider
$d\nu(E)=h(E)dE\in\Mm$ with $h(E)=|E-E'|^{\gamma},\;\;\gamma>-1$, on
an interval around $E'$. Then $\au_{\nu}(E')=1+\gamma$.}
\hfill $\diamond$
\end{rem}

\begin{rem} \label{rem-Lusin} 
{\rm By definition $\gamma<\alpha^-_{\nu}(\Delta)$  if and only if
there exists a set $\Xi\subset\Delta$ of zero $\nu$-measure such that
$\gamma<\alpha_{\nu}(E)$ for all $E\in\Delta\backslash\Xi$.
Furthermore $\gamma<\alpha^+_{\nu}(\Delta)$  if and only if there
exists a set $\Xi\subset\Delta$ of stictly positive $\nu$-measure such
that $\gamma<\alpha_{\nu}(E)$ for all $E\in\Xi$. Because the Borel
function $E\mapsto \int_0^1 \frac{d\epsilon}{\epsilon^{1+\gamma}}
\int_{E-\epsilon}^{E+\epsilon} d\nu(E') $ is bounded on $\Xi$, Lusin's
theorem then implies that there exists a set $\Xi'\subset\Xi$ of
positive $\nu$-measure such that $\int_0^1
\frac{d\epsilon}{\epsilon^{1+\gamma}}  \int_{E-\epsilon}^{E+\epsilon}
d\nu(E') $ has a uniform bound for all $E\in\Xi'$.
}
\hfill $\diamond$
\end{rem}

\begin{defini} {\rm \cite{Str,Com,Las}} The uniform dimension
$\alpha^{\mbox{\rm\tiny uni}}_{\nu}(\Delta)$ of a measure $\nu$ on a
Borel set $\Delta\subset\RR$ is defined by
$$
\alpha^{\mbox{\rm\tiny uni}}_{\nu}(\Delta)\;=\;\sup\{\gamma\in\RR\;|\;
\exists\;C<\infty,\;\delta>0:\;
\int^{E+\epsilon}_{E-\epsilon}\nu(dE')
\;\leq\;C\epsilon^\gamma\;\;
\forall\;\;\epsilon<\delta,\;E\in\Delta\}
\mbox{ . }
$$
\end{defini}

\begin{rem} 
{\rm One clearly has $\alpha^{\mbox{\rm\tiny uni}}_{\nu}(\Delta) \leq
\alpha^-_{\nu}(\Delta)$. However, one does not necessarily have
equality. For if  

$$ f(E) = \sum_{q=2}^{\infty} 
	\sum_{p=1}^{q-1} 
	 \frac{1}{q^2(q-1)}
	  \frac{1}{|E-p/q|^{1/2}}
\mbox{ , }
$$

\noindent then $f\in L^1([0,1])$ and defines an absolutely continuous
probability measure $\nu = z^{-1} f dx$ (if $z>0$ is a normalization factor).
Thus, for any Borel subset $\Delta$ of $[0,1]$,  $\alpha^-_{\nu}(\Delta) =1$
whereas if $\Delta$ contains some rational point, $\alpha^{\mbox{\rm\tiny
uni}}_{\nu}(\Delta) \leq 1/2$.}
\hfill $\diamond$
\end{rem}

\vspace{.2cm}

Now we present some  further technical results as well as
proofs of the other results of Sections \ref{sec-expoess} and 
\ref{sec-expoop}.

\begin{lemma} 
\label{lem-mesura2}
Let $N\in\NN$, $\gamma>0$.
If $\Delta\subset$ is a Borel
set, then 
$$
\Mm^-(\Delta,\gamma,N)\;=\;\{\nu\in\Mm\;|\;
\int^1_0\frac{d\epsilon}{\epsilon^{1+\gamma}}
\;\int^{E+\epsilon}_{E-\epsilon}d\nu(E)
\;\leq \;N \mbox{ for }\nu\mbox{-a.a. }E\in\Delta \}
$$
\noindent and
$$
\Mm^+(\Delta,\gamma,N)\;=\;\{\nu\in\Mm\;|\;
\exists\;\Xi\subset\Delta,\;\nu(\Xi)>0,\; 
\int^1_0\frac{d\epsilon}{\epsilon^{1+\gamma}}
\;\int^{E+\epsilon}_{E-\epsilon}d\nu(E')
\;\leq \;N \mbox{ for }E\in\Xi \}
$$

\noindent are Borel sets in $\Mm$. 
Furthermore, 

$$
\Mm^\pm(\Delta,\gamma,\infty)\;=\;\{\nu\in\Mm\;|\;
\gamma<\alpha^\pm_\nu(\Delta)\}
$$

\noindent are Borel sets.
\end{lemma}

\noindent {\bf Proof.} Let $g_k(x)$ be a continuous non decreasing real
function, equal to $0$ for $x<0$, equal to $1$ for $x>1/k$ and $0\leq g_k(x)\leq
1$ elsewhere. For $\chi\in C_0(\RR)$, $\delta>0$, $N\in\NN$ and $\gamma>0$, the
function 

$$
\nu\in\Mm\;\rightarrow\;
G_{k,\delta,\chi,N,\gamma}(\nu)\;=\;
\int_\RR d\nu(E)\;\chi(E)
\;g_k\left( \int^1_{\delta}
\frac{d\epsilon}{\epsilon^{1+\gamma}}
\nu([E-\epsilon,E+\epsilon]) -N\right)
\mbox{ . }
$$

\noindent is a continuous function. It is non-increasing in $\delta$.
Since $\Delta$ is a Borel set, there exists a sequence
$\chi_{n_1,m_1,\ldots,n_r,m_r}\in C_0(\RR) $, increasing in the $m_j$
and decreasing in the $n_i$, such that the characteristic function
$\chi_\Delta$ is given by $\inf_{n_1}\sup_{m_1}\dots
\inf_{n_r}\sup_{m_r}$ $\chi_{n_1,m_1,\ldots,n_r,m_r}$. Now

$$
G_{\Delta,\gamma,N}(\nu)\;=\;
\inf_{n_1}\sup_{m_1}\dots
\inf_{n_r}\sup_{m_r}\;\sup_k\;\sup_\delta\;
G_{k,\delta,\chi_{n_1,m_1,\ldots,n_r,m_r},N,\gamma}(\nu)
$$

\noindent is a Borel function in $\nu$. By the dominated convergence
theorem

$$
G_{\Delta,\gamma, N}(\nu)\;=\;
\int_\Delta d\nu(E)\;
\;g_{\infty}\left( \int^1_{0}
\frac{d\epsilon}{\epsilon^{1+\gamma}}
\nu([E-\epsilon,E+\epsilon]) -N\right)
\mbox{ . }
$$

\noindent If $G_{\Delta,\gamma,N}(\nu)=0$ then there exists a set
$\Xi$ of $\nu$-measure zero such that $\int^1_{0}
\frac{d\epsilon}{\epsilon^{1+\gamma}} \nu([E-\epsilon,E+\epsilon])
\leq N\;\;\forall\;E\in\Delta\backslash\Xi$. Hence,
$G_{\Delta,\gamma,N}(\nu)=0$ implies  $\nu\in\Mm^-(\Delta,\gamma,N)$.
Clearly $\nu\in\Mm^-(\Delta,\gamma,N)$ implies
$G_{\Delta,\gamma,N}(\nu)=0$. Consequently $\Mm^-(\Delta,\gamma,N)
=G_{\Delta,\gamma,N}^{-1}(\{0\})$ is a Borel set.
$\Mm^+(\Delta,\gamma,N)$ is treated in a similar way. Finally, the last
result follows from $\Mm^\pm(\Delta,\gamma,\infty)=\
\bigcup_{N,M\in\NN}\Mm^\pm(\Delta,\gamma+\frac{1}{M},N)$.
\hfill $\Box$

\vspace{.2cm}

\noindent {\bf Proof} of Proposition \ref{prop-essexpoborel}. Let $h$
denote the application $\nu\in\Mm\mapsto\alpha_\nu^+(\Delta)$. If
$I=(a,b)$ is an open interval, it is sufficient to show that
$h^{-1}(I)$ is a Borel set in order to deduce that $h$ is a Borel
function. With the notations of Lemma \ref{lem-mesura2},

$$
h^{-1}(I) 
\;=\;
\Mm^+(\Delta,a,\infty)\;\cap\;
\left( \bigcap_{n\in\Nn}\Mm^+(\Delta,b-\frac{1}{n},\infty)
\right)^C
\mbox{ , }
$$

\noindent so that $h^{-1}(I)$ is a Borel set by Lemma
\ref{lem-mesura2}. The case of $\alpha_\nu^-(\Delta)$ is treated in a
similar way.
\hfill $\Box$

\vspace{.2cm}

Let $\HH$ be a separable Hilbert space. We consider the
space $\MH$ of $\HH$-projection-valued Borel measures on $\RR$
\cite{RS} endowed with the weak and vague topology, that is (because
of the polarization identity)

$$
\Pi_n\;\stackrel{\MH}{\rightarrow}\;\Pi
\qquad
\Leftrightarrow
\qquad
\int_{\RR}\langle \phi|\Pi_n(dE)|\phi\rangle f(E)\;\rightarrow\;
\int_{\RR}\langle \phi|\Pi(dE)|\phi\rangle f(E)\mbox{ , }
$$

\noindent for all $\phi\in\HH$ and $f$ in $C_0(\RR)$. To every self-adjoint
operator $H$ the spectral theorem  associates a $\Pi\in\MH$.
Convergence in the strong resolvent sense corresponds to convergence
in $\MH$. 

\begin{lemma} 
\label{lem-sepa} 
Let $\Pi\in\MH$ be a $\HH$-projection valued Borel measure on a separable
Hilbert space $\HH$. Then there exists $\psi\in\HH$ so that the
spectral measure $\rho_\psi$ is in the same measure class as $\Pi$.
\end{lemma}

\noindent {\bf Proof.} The lemma being well known, we only sketch an
outline of the proof.
A countable family of normalized vectors $(\phi_i)_{i\in I}$ is
called $\Pi$-free if and only if  $\int\langle
\phi_i|\Pi(dE)|\phi_j\rangle f(E)=0$ for all $i\neq j$ and all $f\in
C_0(\RR)$. The set of $\Pi$-free families is ordered by inclusion and
Zorn's lemma assures the existence of a maximal family $(\phi_i)_{i\in
I}$. Set

$$
\psi\;=\;
c_I
\;\sum_{n\in I}\frac{1}{2^{\frac{n+1}{2}}}\phi_n\mbox{ , }
\qquad
c_I\;=\;\frac{1}{\sqrt{1-2^{-\# I}}}
\mbox{ . }
$$

\noindent It is now possible to verify that the spectral measure
$\rho_\psi$ of $\Pi$ dominates the spectral measure $\rho_\eta$ of any
$\eta\in\HH$.
\hfill $\Box$

\vspace{.2cm}

\noindent {\bf Proof} of Theorem \ref{theo-Piexpogene}. With Lemma
\ref{lem-sepa} choose $\phi\in\HH$  such that $\rho_\psi$ is in the
same measure class as $\Pi$. Then $\rho_\psi$ dominates the spectral
measures $\rho_\phi$ for all $\phi\in\HH$. Hence, by  Theorem
\ref{theo-expoppac},  $\alpha^+_{\rho_\phi}(\Delta)\leq
\alpha^+_{\rho_\psi}(\Delta)\leq
\sup_{\eta\in\HH}\alpha^+_{\rho_\eta}(\Delta)$ and
$\alpha^-_{\rho_\phi}(\Delta)\geq \alpha^-_{\rho_\psi}(\Delta)\geq
\inf_{\eta\in\HH}\alpha^-_{\rho_\eta}(\Delta)$ for all $\phi\in\HH$. 
\hfill $\Box$

\vspace{.1cm}

\begin{proposi}
\label{prop-Piexpogene}
{\bf i)}
\begin{equation}
\label{eq-Piexpo}
\alpha_{\Pi}(E)\;=\;\sup\{\gamma\in\RR\;|\;
\parallel\! \int^1_0\frac{d\epsilon}{\epsilon^{1+\gamma}}
\;\int^{E+\epsilon}_{E-\epsilon}\Pi(dE')\!\parallel_{\Bb(\HH)}
\;<\;\infty\}
\mbox{ . }
\end{equation}

\noindent {\bf ii)}
Let
$G_{\Pi}(z)=\int_{\RR}\frac{\Pi(dE')}{z-E'}$ be the resolvent of
$\Pi$. Suppose $\au_{\Pi}(E)\in [0,2]$ then
$$
\parallel\!\int^1_0\frac{d\epsilon}{\epsilon^{1+\gamma}}
\Im m (G_{\Pi}(E+\imath
\epsilon))\!\parallel_{\Bb(\HH)}\;<\;\infty\mbox{ , }
$$

\noindent if and only if $\gamma<\au_{\Pi}(E)-1$.

\vspace{.1cm}

\noindent {\bf iii)}
$(\Pi,E)\in\MH\times\RR\rightarrow\alpha_{\Pi}(E)$ 
is a Borel function.

\end{proposi}

\begin{proposi}
\label{prop-Piexpogene2}
Let $\Dd$ be a dense subset of
$\HH(\Delta)=\Pi(\Delta)\HH$. Then

$$
\alpha^+_{\Pi}(\Delta)\; = \;\sup_{\phi\in\Dd}
\alpha^+_{\rho_\phi}(\Delta)\mbox{ , }
\qquad
\alpha^-_{\Pi}(\Delta)\; = \;\inf_{\phi\in\Dd}
\alpha^-_{\rho_\phi}(\Delta)\mbox{ . }
$$

\end{proposi}

\noindent The proof of the following lemma follows the lines of the proof of 
Lemma~\ref{lem-mesura2}

\begin{lemma} 
\label{lem-mesura}
Let $N\in\NN$, $\gamma>0$. The set

\begin{equation}
\label{eq-shat}
\hat{\Ss}(\gamma,N)\;=\;\{(\Pi,E)\in\MH\times\RR\;|\;
\forall\;\phi\in\HH:\; \int^1_0\frac{d\epsilon}{\epsilon^{1+\gamma}}
\;\int^{E+\epsilon}_{E-\epsilon}\langle\phi |\Pi(dE')|\phi\rangle
\;\leq\;N\}\mbox{ , }
\end{equation}

\noindent is closed in $\MH\times\RR$. Furthermore, 
$\hat{\Ss}(\gamma,\infty)=\{(\Pi,E)\in\MH\times\RR|\gamma<\alpha_\Pi(E)\}$ 
is a Borel set. 
\end{lemma}

\noindent {\bf Proof} of Proposition \ref{prop-Piexpogene}.
{\bf i)} Let $\beta_\Pi(E)$ be the exponent on the right hand side of
(\ref{eq-Piexpo}). Clearly $\beta_{\Pi}(E)\leq\au_{\Pi}(E)$. To show
$\beta_{\Pi}(E)\geq\au_{\Pi}(E)$, let $\gamma<\au_{\Pi}(E)$. By the
Schwarz inequality, the expression
$$
|\langle \psi|
\int^1_0\frac{d\epsilon}{\epsilon^{1+\gamma}}
\Pi([E-\epsilon,E+\epsilon])
|\phi\rangle |\;\leq\;
\sqrt{\int^1_0\frac{d\epsilon}{\epsilon^{1+\gamma}}
\int_{E-\epsilon}^{E+\epsilon}d\rho_{\psi}(E')}
\sqrt{\int^1_0\frac{d\epsilon}{\epsilon^{1+\gamma}}
\int_{E-\epsilon}^{E+\epsilon}d\rho_{\phi}(E')}
$$

\noindent is bounded for all $\psi,\phi\in\HH$. Consequently, the
positive operator $\int^1_0\frac{d\epsilon}{\epsilon^{1+\gamma}}
\Pi([E-\epsilon,E+\epsilon])$ is everywhere defined. By the
Hellinger-Toeplitz theorem \cite{RS} it is therefore a bounded
operator.  Hence $\gamma<\beta_{\Pi}(E)$.

{\bf ii)} This follows from Theorem \ref{theo-stieltjes} and
an application of the Hellinger-Toeplitz theorem to  

$$
\parallel\!\int_0^{1}\frac{d\epsilon}{\epsilon^{1+(\gamma-1)}}
\Im m (G_{\Pi}(E-\imath \epsilon))
\!\parallel_{\Bb(\HH)}\;=\;
\parallel\!\int_0^{1}\frac{d\epsilon}{\epsilon^{1+\gamma}}
\int_{\RR}\Pi(dE')\;\frac{\epsilon^2}{(E-E')^2+\epsilon^2}
\!\parallel_{\Bb(\HH)}
\mbox{ ,}
$$

\noindent similar to {\bf i)}.

{\bf iii)} 
Let $h$ be  the application $(\Pi,E)\mapsto \au_{\Pi}(E)$. Let
$I=(a,b)$ be an open interval. Then

$$
h^{-1}(I) 
\;=\;
\hat{\Ss}(a,\infty)\;\cap\;
\left(\bigcap_{n\in\NN}\;\hat{\Ss}(b-\frac{1}{n},\infty)\right)^C\mbox{ , }
$$

\noindent and Lemma \ref{lem-mesura} assures that $h^{-1}(I)$ is a
Borel set. Therefore $h$ is a Borel function.
\hfill $\Box$

\vspace{.2cm}

\noindent {\bf Proof} of Proposition \ref{prop-Piexpogene2}.
Let us put $\beta=\sup_{\phi\in\Dd} \alpha^+_{\rho_\phi}(\Delta)$.
Clearly $\beta\leq \alpha^+_{\Pi}(\Delta)$. 

Let now $\psi\in\HH$ be as in Theorem \ref{theo-Piexpogene}
and introduce
$\Xi(\beta)=\{E\in\Delta|\alpha_{\rho_\psi}(E)\leq\beta\}$. As
$E\mapsto\alpha_{\rho_\psi}(E)$ is a Borel function by Theorem
\ref{theo-expoppac}, $\Xi(\beta)$ is a Borel set. Thus
$\HH(\beta)=\Pi(\Xi(\beta))\HH$ is a closed linear subspace of
$\HH(\Delta)$. Now for any $\phi\in\Dd$, $\rho_\psi$ dominates
$\rho_\phi$ and therefore
$\alpha_{\rho_\phi}(E)=\alpha_{\rho_\psi}(E)$ $\rho_\phi$-almost
surely by Theorem \ref{theo-expoppac}. Therefore

$$
\rho_\phi(\Delta)\;\geq\;
\rho_\phi(\Xi(\beta))\;\geq\;
\rho_\phi(\{E\in\Delta\;|\;\alpha_{\rho_\psi}(E)\leq
\alpha^+_{\rho_\phi}(\Delta)\})\;=\;
\rho_\phi(\Delta)
\mbox{ . }
$$

\noindent Hence
$\rho_\phi(\Xi(\beta))=\parallel\!\Pi(\Delta)\phi\!\parallel^2=1$ and
$\phi\in\HH(\beta)$ for all $\phi\in\Dd$. Because $\Dd$ is dense in
$\HH(\Delta)$ by hypothesis,  $\HH(\Delta)=\HH(\beta)$. Consequently,
$\rho_\psi(\Xi(\beta))=\rho_\psi(\Delta)$ and
$\alpha^+_{\rho_\psi}(\Delta)\leq\beta$. Theorem \ref{theo-Piexpogene}
implies $\alpha^+_{\Pi}(\Delta)\leq\beta$. This shows the first
equality. In order to show the second equality, one proceeds in a
similar way using the set of all $E\in\Delta$ such that
$\beta\leq\alpha_{\rho_\psi}(E)$. \hfill $\Box$

\vspace{.2cm}

\subsection{Multifractal dimensions}
\label{sec-global}

The formul{\ae} on which the  multifractal analysis developed below is
based are already explicit in the article of Hentschel and Procaccia
\cite{HP}.  The dimensions  introduced are often referred to as
generalized R\'enyi dimensions \cite{Pes,Ols}. The main reason why
this multifractal analysis is relevant for the quantum-mechanical
study of solids is the following: the behavior of the Fourier
transform of a measure at infinity which is of interest for physicists
\cite{HA1,GK,Hol} can be rigorously linked to the 2-spectral dimension
of the measure, its correlation dimension. This will be done in the
next section. Moreover, the multifractal dimensions give lower bounds
on the lower essential dimension. Note that there are other
possibilities to define multifractal dimensions \cite{HP,Kad,Ols,Pes}.

\begin{defini}
\label{def-multi}
Let $\nu\in\Mm$ and $\Delta\subset\RR$ be a Borel set. If
$\nu(\Delta)\neq 0$, let for $q\in\RR$ 

\begin{equation}
\label{eq-Iq}
I_{\nu}^{q,\epsilon}(\Delta)\;=\;
\lim_{p\downarrow q}\left(
\int_{\Delta}\frac{d\nu(E)}{\nu(\Delta)}
\left(\int_{E-\epsilon}^{E+\epsilon}d\nu(E')
\right)^{p-1}\right)^{\frac{1}{p-1}}\mbox{ , } 
\end{equation}

\noindent The $q$-spectral dimension  $\au_{\nu}^q(\Delta)$ is defined by

$$
I_{\nu}^{q,\epsilon}(\Delta)\;
\stackrel{{\textstyle \sim}}{{\scriptscriptstyle 
\epsilon\downarrow 0}}
\;\epsilon^{\au_{\nu}^q(\Delta)}\mbox{ , }
$$

\noindent unless $I_{\nu}^{q,\epsilon}(\Delta)$ is infinite for a set
of $\epsilon$'s of positive Lebesgue measure {\rm (}possible if
$q\leq 1${\rm )}. We denote  $\au_{\nu}^q=\au_{\nu}^q(\RR)$.  The
dimensions $\au_{\nu}^1$ and $\au_{\nu}^2$ are called information and
correlation dimension respectively.
\end{defini}

\begin{rem} {\rm The notation is chosen such that  in {\sl good} cases
the dimensions $\alpha_{\nu}^q$ are equal to the $D_q$ appearing in
physics literature \cite{HP,Kad,GK}. The dimensions $\alpha_{\nu}^q$
are rigorously linked to box-counting dimensions in \cite{Pes,Ols}. 
}\hfill $\diamond$
\end{rem}

\begin{rem} {\rm  The limit in (\ref{eq-Iq}) is only introduced in
order to study the case $q=1$. Using the monotone convergence theorem
one gets

\begin{equation}
\label{eq-diminfo}
I_{\nu}^{1,\epsilon}(\Delta)\;=\;\mbox{exp}\left(
\int_{\Delta}\frac{d\nu(E)}{\nu(\Delta)}
\;\mbox{Log}\int_{E-\epsilon}^{E+\epsilon}d\nu(E')
\right)
\mbox{ . }
\end{equation}

\noindent This explains why one talks of information dimension. 
}\hfill $\diamond$
\end{rem}

\begin{proposi}
\label{pro-multiexpo}
Let $\nu\in\Mm$  and let $\Delta\subset\RR$ be
a Borel set.

\vspace{.1cm}

\noindent {\bf i)} For $q> 1$, 
$0\leq \au_{\nu}^q(\Delta) \leq 
\alpha_{\nu}^-(\Delta)$.

\vspace{.1cm}

\noindent {\bf ii)} {\rm \cite{Cut}} For $p\leq q$, 
$\au_{\nu}^p(\Delta)\geq\au_{\nu}^q(\Delta)$.

\vspace{.1cm}

\noindent {\bf iii)} $(q-1)\au_{\nu}^q(\Delta)$ is a convex function
of $q$.

\vspace{.1cm}

\noindent {\bf iv)} $\alpha_{\nu}^-(\Delta)\leq
\au_{\nu}^1(\Delta)$.
\end{proposi}

\begin{proposi}
\label{pro-cordim}
Let $\nu\in\Mm$  and $\Delta\subset\RR$
a Borel set.

\vspace{.1cm}

\noindent {\bf i)}
If $\nu \ast\nu$ is the convolution of $\nu$ with
itself, then $\au_{\nu \ast\nu}(0)=\au_{\nu}^2(\RR)$.

\vspace{.1cm}

\noindent {\bf ii)} 
\begin{equation}
\label{eq-poten}
\au_{\nu}^2(\Delta)\;=\; \sup\{\gamma \in \RR\;|\; 
\int_{\Delta} d\nu(E)
\int_{E-1}^{E+1}d\nu (E')\;|E-E'|^{-\gamma}\;<\;\infty\;\}
\mbox{ . }
\end{equation}

\vspace{.1cm}

\noindent {\bf iii)} {\rm \cite{RJLS2}}
The correlation dimension can be calculated as
$$
\int_{\RR}da\;|\Im m G_{\nu}(a+\imath {\epsilon})|^2\;
\stackrel{{\textstyle \sim}}{{\scriptscriptstyle \epsilon\downarrow
0}}
\;\epsilon^{\au_{\nu}^2-1}
\mbox{ . }
$$

\end{proposi}

\begin{rem} \label{rem-mesudep} {\rm  The multifractal dimensions are
not measure class invariants. Let us give an example of an absolutely
continuous measure for which the correlation dimension is smaller than
$1$:

$$
d\nu(E)\;=\;\mbox{const}\frac{1}{E^\beta}e^{-E}\chi(E>0)\mbox{ , }
$$

\noindent where $\chi$ is the indicator function. It is a matter of
calculation to verify that
$\alpha^2_{\nu}=\mbox{min}\{1,2(1-\beta)\}$. This certainly limits
their importance for a mathematical characterization of fractal
measures.
}\hfill $\diamond$
\end{rem}

\noindent {\bf Proof} of Proposition \ref{pro-multiexpo}. {\bf i)}
Since $q>1$, we get $I_{\nu}^{q,\epsilon}(\Delta)\leq 1$ for all
$\epsilon>0$. Hence $\au_{\nu}^q(\Delta)\geq 0$. 

Because $I_{\nu}^{q,\epsilon}(\Delta)$ is increasing in
$\epsilon$, Proposition \ref{prop-diffexpo}v) and Fubini's theorem
shows that

$$
(q-1)\au_{\nu}^q(\Delta)\;=\;\sup\{\gamma\in\RR\;|\;
\int_{\Delta}\frac{d\nu(E)}{\nu(\Delta)}
\int^1_0\frac{d\epsilon}{\epsilon^{1+\gamma}}
\left(\int_{E-\epsilon}^{E+\epsilon}d\nu(E')
\right)^{q-1}<\infty\;\}\mbox{ . }
$$

\noindent Thus, for $\gamma<(q-1)\au_{\nu}^q(\Delta)$ and $\nu$-almost
all $E\in\Delta$, $\int^1_0\frac{d\epsilon}{\epsilon^{1+\gamma}}
\left(\int_{E-\epsilon}^{E+\epsilon}d\nu(E') \right)^{q-1}<\infty$.
Using again the monotonicity in $\epsilon$ and Proposition
\ref{prop-diffexpo}v), $\int^1_0\frac{d\epsilon}{\epsilon^{1+\gamma'}}
\int_{E-\epsilon}^{E+\epsilon}d\nu(E')<\infty$ for all
$\gamma'<\frac{\gamma}{q-1}$ and for $\nu$-almost all $E\in\Delta$,
namely $\gamma'\leq\au_{\nu}(E)$ for $\nu$-almost all $E\in\Delta$.
Hence $\gamma'\leq\alpha_{\nu}^-(\Delta)$ and therefore
$\au_{\nu}^q(\Delta) \leq\alpha_{\nu}^-(\Delta)$.

{\bf ii)} Let us show that for $q\neq 0$,
$I_{\nu}^{q,\epsilon}(\Delta)$ is non-increasing functions of
$\epsilon$ and a  non-decreasing function in the variable $q$. This
implies directly the result. Jensen's inequality for the convex
function $f(t)=t^{\beta}$, $\beta\geq 1$ or $\beta \leq 0$, is used in
the following way: let $p<1<q$ or $1<p<q$, then

$$
\int_{\Delta}\frac{d\nu(E)}{\nu(\Delta)}
\left( \left(\int_{E-\epsilon}^{E+\epsilon}d\nu(E')
\right)^{p-1}\right)^{\frac{q-1}{p-1}}
\; \geq \;
\left(
\int_{\Delta}\frac{d\nu(E)}{\nu(\Delta)}
\left(\int_{E-\epsilon}^{E+\epsilon}d\nu(E')
\right)^{p-1}\right)^{\frac{q-1}{p-1}}
\mbox{ , }
$$

\noindent that is, $I_{\nu}^{q,\epsilon}(\Delta) \leq
I_{\nu}^{p,\epsilon}(\Delta)$. The case $p<q<1$ is treated in a
similar way.

{\bf iii)} Let $q=\sigma q_0+(1-\sigma)q_1$ with
$\sigma\in[0,1]$. By H\"older's inequality one gets

$$
(I_{\nu}^{q,\epsilon}(\Delta))^{q-1}\;\leq\;
(I_{\nu}^{q_0,\epsilon}(\Delta))^{\sigma(q_0-1)}
(I_{\nu}^{q_1,\epsilon}(\Delta))^{(1-\sigma)(q_1-1)}
\mbox{ . }
$$

\noindent Because the right hand side is increasing in $\epsilon$,
Proposition \ref{prop-diffexpo}v) implies that it behaves as
$\sigma(q_0-1)\au_{\nu}^{q_0}(\Delta)
+(1-\sigma)(q_1-1)\au_{\nu}^{q_1}(\Delta)$.  The exponent of the left
hand side is $(q-1)\au_{\nu}^{q}(\Delta)$. The above inequality now
implies that $(q-1)\au_{\nu}^{q}(\Delta)
\geq\sigma(q_0-1)\au_{\nu}^{q_0}(\Delta)
+(1-\sigma)(q_1-1)\au_{\nu}^{q_1}(\Delta)$.

{\bf iv)}  Because $I_{\nu}^{1,\epsilon}(\Delta)$  is
monotone in $\epsilon$, Proposition \ref{prop-diffexpo}iv) and
equation (\ref{eq-diminfo}) imply that the exponent
$\au_{\nu}^1(\Delta)$ is given by

$$
\au_{\nu}^1(\Delta)
\;= \; \liminf_{\epsilon\rightarrow 0}
\int_{\Delta}\frac{d\nu(E)}{\nu(\Delta)}\;
\frac{\mbox{Log}\int_{E-\epsilon}^{E+\epsilon}d\nu(E')}{\mbox{Log }\epsilon}
\mbox{ . }
$$

\noindent By Fatou's lemma and again Proposition
\ref{prop-diffexpo}iv), $\au_{\nu}^1(\Delta) \geq\nu\!-\!
\essinf_{E\in\Delta}\;\au_{\nu}(E)$.
\hfill $\Box$

\vspace{.2cm}

\noindent {\bf Proof} of Proposition \ref{pro-cordim}.
{\bf i)} By definition of the convolution

$$
\int^{E+\epsilon}_{E-\epsilon}d\nu\ast\nu(E')
\;=\;
\int_{\mbox{diag}(E,\epsilon)}d\nu(E')d\nu(E'')
\mbox{ , }
$$

\noindent where diag$(E,\epsilon)=
\{(E',E'')|\;|E-E'+E''|<\epsilon\}$. But for $E=0$, this  last
expression is just equal to $I^{2,\epsilon}_\nu(\RR)$. Therefore
$\au_{\nu \ast\nu}(0)=\au_{\nu}^2$.

{\bf ii)} follows by direct calculation using Fubini's theorem.

{\bf iii)} Because of Theorem \ref{theo-stieltjes} and
$\au_{\nu \ast\nu}(0)=\au_{\nu}^2(\RR)$, it is sufficient to show that

\begin{equation}
\label{eq-linkGreen}
\Im m G_{\nu\ast\nu}(-2\imath {\epsilon})\;=\;\frac{1}{\pi}
\int_{\RR}da\;|\Im m G_{\nu}(a+\imath {\epsilon})|^2
\end{equation}

\noindent For this purpose, write out the right hand side explicitly
and use Fubini's theorem. The contour of the integral over $a$ can be
closed by half of a circle  in the upper half plane because the
integrand falls off as $1/a^4$ at infinity. There are two poles within
the closed contour at $E+\imath\epsilon$ and  $E'+\imath\epsilon$. The
residue theorem then allows to show (\ref{eq-linkGreen}).
\hfill $\Box$

\vspace{.2cm}

\subsection{Asymptotic behavior of Fourier transforms}
\label{sec-CT}

In this section we study the asymptotic behavior of the Fourier
transform of measures on the real line. It is governed by the
correlation dimension of the measure. The first rigorous results in
this direction were obtained by Strichartz \cite{Str}. He gave an
estimate of the decrease of the Fourier transform by the uniform
dimension of the measure (compare Theorem \ref{prop-strich}).
Physicists interest began with the numerical works of \cite{HA1} as
well as \cite{GK}. The latter work also contains a formal derivation
of Theorem \ref{theo-fourier}. Wavelet transform was used in a more
mathematical approach in \cite{Hol}, further related results appear in
\cite{Las,BCM,HoG}. Here, we present two versions of these results as
well as an application to a quantitative version of the RAGE-theorem.
These results are not new and we give them for sake of completeness.

The Fourier transform of $\nu$ is given by
$\Ff_{\nu}(t)=\int d\nu(E)e^{\imath tE}$. Further let

\begin{equation}
\label{eq-CT}
C_{\nu}(T)\;=\;
\int_0^T \frac{dt}{T}\;|\Ff_{\nu}(t)|^2
\mbox{ . }
\end{equation}

\begin{theo}
\label{theo-fourier}
Let $\nu$ be a measure on the real line and $\au^2_{\nu}$ its
correlation dimension. Then
$$
C_{\nu}(T)
\; \stackrel{{\textstyle \sim}}{{\scriptscriptstyle T\uparrow
\infty}} \;T^{-\au^2_{\nu}}
\mbox{ . }
$$

\end{theo}

\noindent The next corollary follows directly from Theorem 
\ref{theo-fourier}, the Schwarz inequality and the equivalent of
Proposition \ref{prop-diffexpo}v) for the behavior at infinity.

\begin{coro}
\label{pro-conv}
Let $\nu,\tilde{\nu}$ be two measures on the real  line. If the
functions $C_{\nu}(T)$ and $C_{\tilde{\nu}}(T)$ are eventually
non-increasing, then

$$
\int_0^T\frac{dt}{T}\;|\Ff_{\nu}(t){\Ff_{\tilde{\nu}}(t)}|
\; \stackrel{{\textstyle \sim}}{{\scriptscriptstyle T\uparrow
\infty}} \;T^{-\alpha}
\qquad \mbox{with }\;\;
\alpha\;\geq\;\frac{1}{2}(\au^2_{\nu}+\au^2_{\tilde{\nu}})\;\geq\;
\min\{\au^2_{\nu},\au^2_{\tilde{\nu}}\}\mbox{ . }
$$

\noindent Moreover, if $\nu\ast\tilde{\nu}$ denotes the  additive
convolution, then one has $\;\;\au^2_{\nu\ast\tilde{\nu}}\;
\geq\;(\au^2_{\nu}+\au^2_{\tilde{\nu}})$.
\end{coro}

\begin{rem} {\rm It is certainly necessary to take the time average in
equation (\ref{eq-CT}). Consider, for example, an absolutely
continuous measure $d\nu(E)=f(E)dE$ with a $C^{\infty}$ function $f$
of compact support. This implies that $|F_{\nu}(t)|^2\sim t^{-2N}$ for
any $N\in\NN$ although the 2-spectral dimension of $\nu$ is
$\au_{\nu}^2=1$. However, taking the time-average, one obtains
$C_{\nu}(T)\sim T^{-1}$ in agreement with Theorem \ref{theo-fourier}.
}\hfill $\diamond$
\end{rem}

\begin{theo}
\label{prop-strich} {\rm \cite{Str}}
Let $\nu\in\Mm$. If $\nu$ has uniform dimension
$\alpha^{\mbox{\rm\tiny uni}}_\nu(\Delta)$ on a Borel set
$\Delta\subset\RR$ and $f\in L^2(\Delta,d\nu)$, then for any
$\beta<\alpha^{\mbox{\rm\tiny uni}}_\nu(\Delta)$ and some positive
constant $C$:

\begin{equation}
\label{eq-defbeta}
\int_0^T\frac{dt}{T}
|\int_{\Delta}d\nu(E)\;f(E)\;e^{\imath Et}|^2
\;\leq\;C\parallel\! f\!\parallel^2_{L^2(\Delta,d\nu)}
 \;T^{-\beta}
\mbox{ . }
\end{equation}
\end{theo}

\noindent Before coming to the proofs, let us study the links between
the diffusion of presence probability under quantum mechanical time
evolution and spectral properties of the underlying Hamiltonian. This
leads to a quantitative version of the RAGE-theorem \cite{RS}.

\begin{coro} 
\label{coro-RAGE}
Let $\phi$, $\parallel\!\phi\!\parallel=1$,  belong to a Hilbert space
$\HH$. Let $H$  a be self-adjoint operator on $\HH$.  Let
$\rho_{\phi}$ be the spectral  measure of $H$ associated to $\phi$.
Then

\begin{equation}
\label{def-exRAGE} 
\int_0^T\frac{dt}{T}\;|\langle \phi|e^{-\imath Ht}|\phi\rangle |^2
\; \stackrel{{\textstyle \sim}}{{\scriptscriptstyle T\uparrow
\infty}} \;T^{-\alpha^2_{\rho_\phi}} \mbox{ . }
\end{equation}

\end{coro}

\begin{rem} {\rm Let $\phi,\psi\in\HH$,
$\parallel\!\phi\!\parallel=1$, $\parallel\!\psi\!\parallel=1$. 
If $\beta$ is defined by

$$
\int_0^T\frac{dt}{T}\;|\langle \phi|e^{-\imath Ht}|\psi\rangle |^2
\; \stackrel{{\textstyle \sim}}{{\scriptscriptstyle T\uparrow
\infty}} \;T^{-\beta} \mbox{ , }
$$

\noindent then

\begin{equation}
\label{def-exRAGE2} 
\beta\;=\;\sup \;\{\gamma\in\RR\;|\;\int_{\RR}d\rho_{\phi,\psi}(E)
\int_{\RR}d\rho_{\psi,\phi}(E')\;|E-E'|^{-\gamma}\;<\;\infty\;\} \mbox{
.}
\end{equation}

\noindent where $\rho_{\psi,\phi}$ is the complex spectral measure of
$H$ associated to $\phi,\psi$. If either $\rho_{\phi}$ or
$\rho_{\psi}$ has uniform dimension $\alpha^{\mbox{\rm\tiny uni}}$,
then one has $\beta\geq\alpha^{\mbox{\rm\tiny uni}}$ as shows directly
the spectral theorem and Theorem \ref{prop-strich}.
}
\hfill $\diamond$
\end{rem}

\vspace{.2cm}

\noindent {\bf Proof} of Theorem \ref{theo-fourier}. First  note that
$0\leq\au^2_{\nu}\leq 1$. We have to show the equivalence between

\begin{equation}
\label{eq-condi1}
\int_{E\geq E'}\frac{d\,\nu(E)d\nu(E')}{|E-E'|^{\gamma}}
\;<\;\infty
\end{equation}

\noindent and

\begin{equation}
\label{eq-condi2}
\int_1^{\infty}\frac{dT}{T^{1-\gamma}}
\int_{E\geq E'}d\,\nu(E)d\nu(E')\;\frac{2\sin((E-E')T)}{(E-E')T}\;
<\;\infty\;\mbox{ , } 
\end{equation}

\noindent provided $0<\gamma<1$. Suppose that (\ref{eq-condi1}) holds.
Because the integral

\begin{equation}
\label{eq-sinest}
\int_1^{\infty}\frac{dT}{T}\;\frac{\sin(|E-E'|T)}{(|E-E'|T)^{1-\gamma}}
\;=\;\int^{\infty}_{|E-E'|}ds\;\frac{\sin s}{s^{2-\gamma}}
\end{equation}

\noindent is absolutely convergent and behaves like $\Oo(1)$ as
$|E-E'|\rightarrow 0$, we may apply Fubini's theorem to 

$$
\int_{E\geq E'}\frac{d\,\nu(E)d\nu(E')}{|E-E'|^{\gamma}}\;
\int_1^{\infty}\frac{dT}{T}\;\frac{\sin((E-E')T)}
{((E-E')T)^{1-\gamma}} \;<\infty
$$

\noindent in order to deduce (\ref{eq-condi2}).

For the converse, let $N$ be any integer. By
(\ref{eq-condi2}), there is a constant $C$ independent of $N$ such
that 
{\small
$$
C\;>
 \int_1^N \frac{dT}{T^{1-\gamma}}
  \int_{E-E'\geq 0} d\,\nu(E)d\nu(E')
   \frac{\sin((E-E')T)}{(E-E')T}
>
\int_{E-E'\geq \frac{\pi}{N}}
 \frac{d\,\nu(E)d\nu(E')}{|E-E'|^{\gamma}}
  \int_{E-E'}^{N(E-E')} ds\;
   \frac{\sin s}{s^{2-\gamma}}
\mbox{ , }
$$
}
\noindent because the integrand is positive as long as $E-E'\;<\;
\pi/N$. We have used Fubini's theorem. Then 
$0 < \int^\infty_0ds \sin(s)/s^{2-\gamma}< \infty$ implies (\ref{eq-condi1}).
\hfill $\Box$

\vspace{.2cm}

\noindent {\bf Proof} of Theorem \ref{prop-strich}. Use the Cauchy-Schwarz
inequality and $|\sin(\theta)|\leq |\theta|^{1-\beta}$ for all $\beta\in[0,1]$,
to get :

$$
\int_0^T\frac{dt}{T}
|\int_{\Delta}d\nu(E)\;f(E)\;e^{\imath Et}|^2
\leq\;
\int_{\Delta}d\nu(E)\;|f(E)|^2 \int_{\Delta}d\nu(E')
\left(\frac{2}{|E-E'|T}\right)^\beta
\mbox{ . }
$$

\noindent Now if $\beta<\alpha^{\mbox{\rm\tiny uni}}_\nu(\Delta)$, by
definition there exists  constants $C<\infty$ and $\delta>0$ such that
$\nu((E-\epsilon,E+\epsilon))\leq C\epsilon^\beta$ for all $\epsilon
\leq \delta$. Hence for
$\beta<\beta'<
\alpha^{\mbox{\rm\tiny uni}}_\nu(\Delta)$ one has
(with changing constants $C$):

$$
\int_\RR d \nu(E')\frac{1}{|E-E'|^\beta}\;\leq\;
\beta\int_{E-\delta}^{E+\delta} d \nu(E')
\int^\delta_{|E-E'|}\frac{d\epsilon}{\epsilon^{1+\beta}}
+C
\;\leq\;\beta
\int^\delta_0\frac{d\epsilon}{\epsilon^{1+\beta}}C\epsilon^{\beta'}
+C\;\leq\;C
\mbox{ . }
$$

\noindent As this bound is uniform in $E$, this finishes the proof.
\hfill $\Box$
 
\vspace{.5cm}

\section{Spectral exponents and anomalous quantum diffusion}
\label{chap-expon}

\subsection{Spectral exponents of covariant Hamiltonians}
\label{sec-DOS}

Let $H=H^*\in \Aa$ be a given covariant Hamiltonian family. It gives rise to a
covariant family $(\Pi_{\omega})_{\omega\in\Omega}$ of projection-valued
measures on $\HH=\ell^2(\ZZ^d)$ by

\begin{equation}
\label{eq-defpiome}
\int_{\RR}\Pi_{\omega}(dE)\;f(E)\;=\;\pi_{\omega}(f(H))\mbox{ , }
\qquad f\in C_0(\RR)\mbox{ . }
\end{equation}

\noindent For $\phi\in\HH$, the corresponding spectral measure is
denoted by  $\rho_{\omega,\phi}$. Let us introduce the measure

$$
\rho_\omega\;=\;\sum_{n\in\ZZ^d}c_n
\rho_{\omega,|n\rangle}\mbox{ , }
$$

\noindent where the sequence $(c_n)_{n\in\ZZ^d}$ of positive numbers
satisfies $\sum_{n\in\ZZ^d}c_n=1$. By the following lemma, 
$\rho_\omega$ is in the same  measure class as  $\Pi_\omega$ and the
application $\Pi_\omega\in\MH\mapsto \rho_\omega\in\Mm$ is continuous
for fixed $(c_n)_{n\in\ZZ^d}$.

\vspace{.2cm}

\begin{lemma} 
\label{lem-equicont}
Let $(\phi_n)_{n\in\NN}$ be  an orthonormal basis of $\HH$. Let
$\Pi\in\MH$ and introduce  the measure $\rho_\Pi=\sum_{n=0}^{\infty}
\frac{1}{2^{n+1}}\rho_{\phi_n}$. Then $\rho_\Pi$ is in the same
measure class as $\Pi$ and the application
$\Pi\in\MH\mapsto\rho_\Pi\in\Mm$ is continuous.
\end{lemma}

\noindent {\bf Proof.} If a Borel set $\Delta\subset\RR$ satisfies
$\rho_\Pi(\Delta)=0$, then $\rho_{\phi_n}(\Delta)=0$ for all $n\in\NN$. Hence
for $\psi=\sum_{n\in\NN} a_n \phi_n$, $\parallel\!\psi\!\parallel=1$, the
Schwarz inequality gives $\rho_\psi(\Delta)=0$. Therefore $\rho_\Pi$ dominates
$\Pi$. On the other hand, $\parallel\!\Pi(\Delta)\!\parallel=0$ clearly implies
$\rho_\Pi(\Delta)=0$. Hence $\rho_\Pi$ and $\Pi$ are in the same measure class.
Let now $\Pi_l\rightarrow\Pi$ in $\MH$ as $l\rightarrow\infty$. For $f\in
C_0(\RR)$, by the dominated convergence theorem

$$
\lim_{l\rightarrow\infty}\int d\rho_{\Pi_l}(E)\;f(E)
\;=\;
\sum_{n\in\NN}\frac{1}{2^{n+1}}
\int \langle\phi_n|\Pi(dE)|\phi_n\rangle\;f(E)
\;=\;\int d\rho_{\Pi}(E)\;f(E)
\mbox{ . }
$$

\noindent Therefore, $\rho_{\Pi_l}\rightarrow\rho_{\Pi}$ in $\Mm$ as
$l\rightarrow\infty$. \hfill $\Box$

\vspace{.2cm}

\begin{lemma}
\label{lem-dimLDS}
The applications $\omega\in\Omega \mapsto\Pi_\omega\in\MH$ and
$\omega\in\Omega \mapsto\rho_\omega\in\Mm$ are continuous.
\end{lemma}

\noindent {\bf Proof.}  For $f\in C_0(\RR)$, $f(H)\in\Aa$ and  the
application $\omega \mapsto \pi_{\omega}(f(H))$ is strongly
continuous. Because of equation (\ref{eq-defpiome}) and the definition
of the topology on $\MH$, this implies the continuity of the
application $\omega\in\Omega \mapsto\Pi_\omega\in\MH$. The second
statement now follows from Lemma \ref{lem-equicont}. 
\hfill $\Box$

\vspace{.2cm}

\noindent {\bf Proof} of Theorem \ref{theo-dimLDS}. We denote the
spectral projection $\int_\Delta \Pi_\omega(dE)$ by
$\Pi_\omega(\Delta)$ and  introduce the function

$$
F_{\omega,E}(\gamma)\;=\;\parallel\!
\int^1_0\frac{d\epsilon}{\epsilon^{1+\gamma}}\;
\Pi_{\omega}([E-\epsilon,E+\epsilon])\!\parallel 
\mbox{ . }
$$

\noindent The covariance implies  that
$F_{\omega,E}(\gamma)=F_{T^{-n}\omega,E}(\gamma)$ for all $n\in\ZZ^d$. The sets
$\Omega_{\gamma,N}= \{\omega\in\Omega|F_{\omega,E}(\gamma)\leq N\}$ ($n\in
\NN\cup \{ \infty\} $) are therefore $T$-invariant. 
Moreover,  $\Omega_{\gamma,N}=
\{\omega\in\Omega|(\Pi_{\omega},E)\in\hat{\Ss}(\gamma,N)\}$ is Borel (see Lemma
\ref{lem-mesura}). 
By the ergodicity of $\PP$ $\PP(\Omega_{\gamma,N}) =0$ or $1$. 
The monotonicity of $F_{\omega,E}(\gamma)$ in $\gamma$ implies that, for
$\gamma<\gamma'$, $\Omega_{\gamma',\infty}\subset\Omega_{\gamma,\infty}$. 
Hence there exists a $\gamma_c$ such that for $\gamma<\gamma_c$, $\PP
(\Omega_{\gamma,\infty})=1$, and for $\gamma>\gamma_c$,
$\PP(\Omega_{\gamma,\infty})=0$. Because $\Omega_{\gamma,\infty}=
\{\omega\in\Omega|\gamma<\alpha_{\Pi_\omega}(E)\}$,
$\gamma_c=\alpha_{\Pi_\omega}(E)$ $\PP$-almost surely. 

Let us now consider $\alpha^\pm_{\Pi_\omega}(\Delta)$. The
set $\Omega^\pm_{\gamma,\Delta}=
\{\omega\in\Omega|\gamma<\alpha^\pm_{\rho_\omega}(\Delta)\}$ is
$T$-invariant by Corollary \ref{coro-essexpo} because $\rho_{\omega}$
and $\rho_{T^a\omega}$ are in the same measure class for any
$a\in\ZZ^d$. 
Moreover, if $\Mm^\pm(\gamma,\Delta,\infty)$ are  the
sets  defined in Lemma \ref{lem-mesura2}, then
$\Omega^\pm_{\gamma,\Delta}=
\{\omega\in\Omega|\rho_{\omega}\in\Mm^\pm(\gamma,\Delta,\infty)\}$.
The Lemma \ref{lem-mesura2} and \ref{lem-dimLDS} imply that
$\Omega^\pm_{\gamma,\Delta}$ are Borel sets. By ergodicity of $\PP$, they have
either full or zero $\PP$-measure. As $\gamma<\gamma'$ implies
$\Omega^\pm_{\gamma',\Delta}\subset\Omega^\pm_{\gamma,\Delta}$, there
exist critical values $\gamma_c^\pm$ such that
$\PP(\Omega^\pm_{\gamma',\Delta})=1$ for $\gamma'<\gamma^\pm_c$ and
$\PP(\Omega^\pm_{\gamma,\Delta})=0$ for $\gamma_c^\pm<\gamma$.
\hfill $\Box$

\vspace{.2cm}

\noindent {\bf Proof} of Theorem \ref{theo-LDSDOS}. Let
$\gamma<\alpha_{\mbox{\rm\tiny LDOS}}(E)$. Then it follows from the
arguments in the  previous proof of Theorem \ref{theo-dimLDS} that
there exists an  $N<\infty$ such that the $T$-invariant set
$\{\omega\in\Omega|(\Pi_{\omega},E)\in\Ss(\gamma,N)\}$ has full
measure. On this set of full measure the $F_{\omega,E}(\gamma)$'s have
uniform bound $N$, $\PP$-almost surely. Therefore the spectral
exponent satisfies

\begin{eqnarray}
\alpha_{\mbox{\rm\tiny LDOS}}(E) & = & 
\sup\{\gamma\in\RR\;|\;
\int_{\Omega}d\PP(\omega)\;F_{\omega,E}(\gamma)
\;<\infty\}
\nonumber
\\
& \leq &
\sup\{\gamma\in\RR\;|\;
\parallel\!\int_{\Omega}d\PP(\omega)\;\int^1_0
\frac{d\epsilon}{\epsilon^{1+\gamma}}
\int_{E-\epsilon}^{E+\epsilon}\Pi_{\omega}(dE')
\!\parallel
\;<\infty\}\mbox{ , }
\nonumber
\end{eqnarray}

\noindent such that $\alpha_{\mbox{\rm\tiny LDOS}}(E)\leq
\inf_{\phi\in\HH}\;\alpha_{\Nn_{\phi}}(E)$. Thus,
$\alpha_{\mbox{\rm\tiny LDOS}}(E)\leq\alpha_{\mbox{\rm\tiny DOS}}(E)$.

Let now $\Omega_1=\{\omega\in\Omega|
\alpha^\pm_{\rho_\omega}(\Delta)=\alpha^\pm_{\mbox{\rm\tiny
LDOS}}(\Delta) \}$. If $g$ is the continuous application
$\omega\in\Omega \mapsto\rho_\omega\in\Mm$ (by Lemma \ref{lem-dimLDS})
and $h$ the Borel function $\rho_\omega\in\Mm
\mapsto\alpha^\pm_{\rho_\omega}(\Delta)$ (by Proposition
\ref{prop-essexpoborel}), then  $\Omega_1=g^{-1}(h^{-1}(\{
\alpha^\pm_{\mbox{\rm\tiny LDOS}}(\Delta)\})$ is a Borel set. Moreover
$\Omega_1$ is $T$-invariant and has full $\PP$ measure.

Now let $I_{a}=\{(\omega,E)\in\Omega\times\RR|
\alpha_{\rho_{T^a\omega}}(E)=\alpha_{\rho_\omega}(E)\}$ for
$a\in\ZZ^d$. If $k$ is the Borel function
$(\omega,E)\in\Omega\times\RR\mapsto
\alpha_{\rho_{T^a\omega}}(E)-\alpha_{\rho_\omega}(E)$ (by Lemma
\ref{lem-dimLDS} and Theorem \ref{theo-expoppac}iv)), then
$I_{a}=k^{-1}(\{0\})$ shows that $I_{a}$ is a Borel set. Hence
$I=\bigcap_{a\in\ZZ^d}I_{a}$ is also a Borel set. Because
$\rho_{\omega}$ and $\rho_{T^a\omega}$ are in the same measure class,
Theorem \ref{theo-expoppac} gives $\rho_\omega(\{E\in\RR|(\omega,E)\in
I_{a}\})=1$ and by $\sigma$-additivity, $\rho_\omega(\{E\in\RR
|(\omega,E)\in I\})=1$.

Finally, let
$\overline{\Delta}=\{(\omega,E)\in\Omega\times\Delta|
\alpha_{\mbox{\rm\tiny LDOS}}^-(\Delta)
\leq\alpha_{\rho_\omega}(E)\leq \alpha_{\mbox{\rm\tiny
LDOS}}^+(\Delta)\}$. By Lemma \ref{lem-dimLDS} and Theorem
\ref{theo-expoppac}iv),  $\overline{\Delta}$ is a Borel set. It also
satisfies $\rho_\omega(\{E\in\Delta|(\omega,E)\in\overline{\Delta}\})
=\rho_\omega(\Delta)$ for $\PP$-almost all $\omega\in\Omega$. 

Now we set $\hat{\Delta}=I\cap(\Omega_1\times\RR)\cap
\overline{\Delta}$. It is a Borel set and  $\rho_\omega
(\{E\in\Delta|(\omega,E)\in\hat{\Delta}\}) =\rho_\omega(\Delta)$ for
$\PP$-almost all $\omega\in\Omega$.  If $(\omega,E)\in\hat{\Delta}$,
then $(T^a\omega,E)\in\hat{\Delta}$ for all $a\in\ZZ^d$. If
$\chi_{\hat{\Delta}}$ is the indicator function of $\hat{\Delta}$,
then the definition of $\hat{\Delta}$ and Fubini's theorem give

$$
\int d\PP(\omega)\int d\rho_\omega(E)\;\chi_{\hat{\Delta}}(\omega,E)
\;=\;\Nn(\Delta)
\mbox{ . }
$$

On the other hand, the invariance of $\PP$,
$\rho_{\omega,|n\rangle}=\rho_{T^a\omega,|n-a\rangle}$ and Fubini's
theorem imply that

$$
\int d\PP(\omega)\int d\rho_\omega(E)\;\chi_{\hat{\Delta}}(\omega,E)
\;=\;\sum_{n\in\NN}c_n\int d\PP(\omega)
\left(\frac{1}{|\Lambda|}\sum_{m\in\Lambda}
\int d\rho_{\omega,|n-m\rangle}(E)\right)\chi_{\hat{\Delta}}(\omega,E)
\mbox{ . }
$$ 

\noindent for any $\Lambda\subset\ZZ^d$. By Birkhoff's theorem,  in
the limit of increasing rectangles centered at the origin
$\Lambda\rightarrow\ZZ^d$, the term in the parenthesis converges to
$\Nn$, $\PP$-almost surely. Therefore, if we introduce the
$T$-invariant Borel set $\Omega_E=\{\omega\in\Omega
|(\omega,E)\in\hat{\Delta}\}$, Fubini's theorem gives

$$
\Nn(\Delta)\;=\;\int_\Delta d\Nn(E) \;\PP(\Omega_E)
\mbox{ . }
$$

\noindent Hence $\PP(\Omega_E)=1$ for $\Nn$-almost all $E\in\Delta$.
Consequently, because

$$
\Omega_E\;=\;\{\omega\in\Omega|
\alpha_{\mbox{\rm\tiny LDOS}}^-(\Delta)
\leq\alpha_{\rho_\omega}(E)=\alpha_{\rho_{T^a\omega}}(E)\leq
\alpha_{\mbox{\rm\tiny LDOS}}^+(\Delta) \;\;\forall\;a\in\ZZ^d\}
$$

\noindent by definition of $\hat{\Delta}$,  $\alpha_{\mbox{\rm\tiny
LDOS}}^-(\Delta) \leq\alpha_{\mbox{\rm\tiny LDOS}}(E)\leq
\alpha_{\mbox{\rm\tiny LDOS}}^+(\Delta)$ for $\Nn$-almost all
$E\in\Delta$. Because $\alpha_{\mbox{\rm\tiny LDOS}}(E)\leq
\alpha_{\mbox{\rm\tiny DOS}}(E)$, $\alpha_{\mbox{\rm\tiny
LDOS}}^-(\Delta)\leq \alpha_{\mbox{\rm\tiny DOS}}^-(\Delta)$ follows.

In order to show $\alpha_{\mbox{\rm\tiny
LDOS}}^+(\Delta)\leq \alpha_{\mbox{\rm\tiny DOS}}^+(\Delta)$, it is
now sufficient to show that

$$
\Nn\!-\!\esssup_{E\in\Delta}\alpha_{\mbox{\rm\tiny LDOS}}(E)
=\alpha_{\mbox{\rm\tiny LDOS}}^+(\Delta)
\mbox{ . }
$$ 

\noindent For this purpose, fix $\delta>0$ and introduce

$$
\Xi\;=\;I\;\cap\;(\Omega_1\times\RR)\;\cap\;
\{(\omega,E)\in\Omega\times\Delta\;|\;
\alpha_{\mbox{\rm\tiny LDOS}}^+(\Delta)-\delta
\leq\alpha_{\rho_\omega}(E)\leq
\alpha_{\mbox{\rm\tiny LDOS}}^+(\Delta)\}
\mbox{ . }
$$

\noindent Then $\rho_\omega(\{E\in\RR| (\omega,E)\in\Xi\})>0$,
$\PP$-almost surely. Now introduce
$\tilde{\Omega}_E=\{\omega\in\Omega|(\omega,E)\in\Xi\}$. Repeating the
same arguments as above, there exists a set $\Xi_\Nn$ of positive
$\Nn$-measure such that the $T$-invariant Borel set
$\tilde{\Omega}_E$ has full $\PP$-measure, that is for all
$E\in\Xi_\Nn$,

$$
\alpha_{\mbox{\rm\tiny LDOS}}^+(\Delta)-\delta
\;\leq\;\alpha_{\mbox{\rm\tiny LDOS}}(E)\;\leq\;
\alpha_{\mbox{\rm\tiny LDOS}}^+(\Delta)
\mbox{ . }
$$

\noindent As $\delta>0$ is arbitrary, this finishes the proof.
\hfill $\Box$

\vspace{.2cm}

\begin{rem} {\rm  Usually one expects averaged exponents to be smaller
than exponents obtained by taking an essential infimum over disorder
configurations. Actually, it is easy to check that
$\alpha_{\mbox{\rm\tiny DOS}}(E)\leq \PP\!-\!\essinf_{\omega\in\Omega}
\alpha_{\rho_{\omega,|0\rangle }}(E)$. On the other hand, it is
possible that the inequality

$$
\PP\!-\!\essinf_{\omega\in\Omega}
\alpha^-_{\rho_{\omega,|0\rangle }}(\Delta)
\;<\;\alpha^-_{\mbox{\rm\tiny DOS}}(\Delta)
$$

\noindent be realized. This is at the basis of Theorem
\ref{theo-dimLDS}. To understand the difficulty, let us consider a
Hamiltonian with dense pure-point spectrum for $\PP$-almost all
$\omega\in\Omega$. Therefore $\PP\!-\!\essinf_{\omega\in\Omega}
\alpha^-_{\rho_{\omega,|0\rangle }}(\Delta)=0$. It is however well
known that a given $E$ is $\PP$-almost surely not in the spectrum so
that $\alpha_{\mbox{\rm\tiny DOS}}(E)$ may be strictly bigger  than
$0$. The same may then hold for $\alpha^-_{\mbox{\rm\tiny
DOS}}(\Delta)$.
}\hfill $\diamond$
\end{rem}

As the exponents of the LDOS and the diffusion exponent, the
exponents of the DOS do not depend on a given vector in Hilbert space
as  suggests the definition (\ref{eq-DOSdef}). For $\phi\in\HH$,
$\Nn_\phi$ is defined by

$$
\int d\Nn_\phi(E)\;f(E)\;=\;\int d\PP(\omega)\langle
\phi|\pi_\omega(f(H))|\phi\rangle
\mbox{ , }
\qquad
f\in C_0(\RR)\mbox{ . }
$$

\noindent The DOS is then $\Nn=\Nn_{|0\rangle}$. 

\begin{proposi}
\label{prop-dimDOS} For any $\phi\in\HH$, the measure $\Nn_\phi$ is
dominated by $\Nn$. 
If $E\in\RR$ and $\Delta\subset\RR$ a Borel set, then

$$
\alpha_{\mbox{\rm\tiny DOS}}
(E)\;=\;\inf_{\phi\in\HH}\alpha_{\Nn_\phi}(E)
\mbox{ , }
\qquad
\alpha^+_{\mbox{\rm\tiny DOS}}(\Delta)\;=\;
\sup_{\phi\in\HH}\alpha^+_{\Nn_\phi}(\Delta)
\mbox{ , }
\qquad
\alpha^-_{\mbox{\rm\tiny DOS}}(\Delta)\;=\;
\inf_{\phi\in\HH}\alpha^-_{\Nn_\phi}(\Delta)
\mbox{ . }
$$  
\end{proposi}

\noindent {\bf Proof.} By Theorem \ref{theo-expoppac} and Corollary
\ref{coro-essexpo}, it is sufficient to show that $\Nn$ dominates
$\Nn_\phi$ for any $\phi\in\HH$. Let $\Delta\subset\RR$ be such that
$\Nn(\Delta)=0$. If $\phi=\sum_{n\in\ZZ^d}\phi_n|n\rangle \;\in\HH$,
then $A=\sum_{n} {\phi_n}U(n)\in L^2(\Aa,\TV)$ and

$$
\int
d\Nn_\phi(E)\;f(E)\;=\;\TV(f(H)AA^*)\;=\;\langle A|f(H)|A
\rangle_{L^2(\Aa,\TV)}
\mbox{ . }
$$

\noindent Let $A_n\in\Aa$ be a Cauchy sequence in $L^2(\Aa,\TV)$
converging to $A$ such that $A_n$ be a trigonometric polynomial.  Then
$\langle A_n|f(H)|A_n\rangle$ converges to  $\langle A|f(H)|A\rangle$
for any bounded $f(H)$. Since $A_n$ is a trigonometric polynomial,
$\langle A_n|\chi_\Delta(H)|A_n\rangle=0$ where $\chi_\Delta$ is the
characteristic function of the Borel set $\Delta$. Hence $\langle
A|\chi_\Delta(H)|A\rangle=0$, that is $\Nn_\phi(\Delta)=0$.
\hfill $\Box$

\vspace{.2cm}

\subsection{Diffusion exponents of covariant Hamiltonians}
\label{sec-transexpo}

Let us first generalize Definition \ref{def-exdiff}.

\begin{defini}
\label{def-exdiff2}
Let $\delta X^2_{\omega,\Delta}(T)$ be the mean square displacement
operator defined in {\rm (\ref{def-Lsqdisproj})}. For $\phi\in\HH$, 

$$
\int_{\Omega}d\PP(\omega)\;\langle \phi|\delta
X^2_{\omega,\Delta}(T)|\phi\rangle \;
\stackrel{{\textstyle \sim}}{{\scriptscriptstyle T\uparrow \infty}} 
\;T^{2\sigma_{\phi}(\Delta)}
$$
\noindent and
$$
\langle \phi|\delta X^2_{\omega,\Delta}(T)|\phi\rangle 
\;\stackrel{{\textstyle \sim}}{{\scriptscriptstyle T\uparrow \infty}} 
\;T^{2\sigma_{\omega,\phi}(\Delta)}
$$

\noindent define the diffusion exponents $\sigma_{\phi}(\Delta)$ and
$\sigma_{\omega,\phi}(\Delta)$. We set
$\hat{\sigma}_{\phi}(\Delta)\;=\;\PP\!-\!\esssup_{\omega\in\Omega}
\sigma_{\omega,\phi}(\Delta)$ and $\sigma_{\mbox{\rm\tiny
diff}}(\Delta)=\sigma_{|0\rangle }(\Delta)$ where $|0\rangle $ is the
state localized at the origin. 
\end{defini}

\begin{rem} \label{rem-difffluc}
{\rm Clearly $\hat{\sigma}_{\phi}(\Delta)\leq{\sigma}_{\phi}(\Delta)$.
A strict inequality may be possible if  $\langle \phi|\delta
X^2_{\omega,\Delta}(T)|\phi\rangle $ has large fluctuations in
$\omega$.
}
\hfill $\diamond$
\end{rem}

\begin{proposi}
\label{pro-diffgener}
{\bf i)} For any $\phi\in\ell^1(\ZZ^d)$,   $\sigma_{\phi}(\Delta)\leq
\sigma_{\mbox{\rm\tiny diff}}(\Delta)$.

\vspace{.1cm}

\noindent {\bf ii)} If $\phi\in\HH$ satisfies  $\langle
\phi|\Pi_\omega(\Delta)X^2\Pi_\omega(\Delta)|\phi\rangle <\infty$,
then its diffusion exponent $\sigma_{\omega,\phi}(\Delta)$ can also be
calculated by replacing the operator $(\vec{X}_{\omega}(t)-\vec{X})^2$
in {\rm (\ref{def-Lsqdisproj})} by $\vec{X}_{\omega}^2(t)$.

\end{proposi}

The rest of this section is devoted to proofs.

\vspace{.2cm}

\noindent {\bf Proof} of Proposition \ref{prop-exdiff3}. By DuHamel's
formula,

$$
\vec{\nabla}(e^{-\imath Ht})\;=
 -\imath 
  \;\int^t_0 ds\;
   e^{-\imath H(s-t)}\vec{\nabla}(H)e^{-\imath Hs}
\mbox{ , }
$$

\noindent where the integral is defined as a norm-convergent Riemann
sum. Because $H\in\Cc^1(\Aa)$, $\vec{\nabla}(e^{-\imath Ht})$ is an
element of $\Aa$ and hence $\TV(\Pi(\Delta)|\vec{\nabla}(e^{-\imath
Ht})|^2)$ is well defined. Using the definition of the gradient, the
cyclicity of the trace $\TV$ and $[\Pi(\Delta),e^{-\imath Ht}]=0$, one
gets

$$
\TV(\Pi(\Delta)|\vec{\nabla}(e^{-\imath Ht})|^2)
\;=\;
\TV(\Pi(\Delta)e^{\imath Ht}
[\XV,e^{-\imath Ht}]\cdot
[e^{\imath Ht},\XV]e^{-\imath Ht})\;
=\;\TV(\Pi(\Delta)(\XV(t)-\XV)^2\Pi(\Delta))
\mbox{ .}
$$

\noindent By definition of the trace $\TV$ on $L^\infty(\Aa,\TV)$,

$$
\int^T_0\frac{dt}{T}\TV(\Pi(\Delta)(\XV(t)-\XV)^2\Pi(\Delta))
\;=\;
\int_\Omega d\PP(\omega)\;\langle 0|\delta X^2_{\omega,\Delta}(T)|
0\rangle
\mbox{ . }
$$

\noindent This finishes the proof. 
\hfill $\Box$

\vspace{.2cm}

\noindent {\bf Proof} of Theorem \ref{theo-diffgener}i) and ii). {\bf
i)}  $\sigma_{\mbox{\rm\tiny diff}}(\Delta)$ is clearly bigger than or
equal to $0$. Because $H\in\Cc^1(\Aa)$, DuHamel's formula implies

$$
\parallel\! \vec{\nabla} (e^{\imath Ht})\!\parallel\;\leq\;\int_0^t ds
\parallel\! e^{\imath H(t-s)} \vec{\nabla}H
 e^{\imath Hs}\!\parallel
\;\leq\;t\parallel\!  \vec{\nabla}(H)\!\parallel
\mbox{ . }
$$

\noindent so that

$$
\int^T_0\frac{dt}{T}
\TV(\Pi(\Delta)|\vec{\nabla}(e^{-\imath Ht})|^2)
\;\leq\;
\int_0^T\frac{dt}{T}\;\parallel\!
\vec{\nabla} (e^{\imath Ht})\!\parallel^2
\;\leq\;
\frac{1}{3}\parallel\!
\vec{\nabla} (H)\!\parallel^2\,T^2
\mbox{ , }
$$

\noindent Hence the exponent  $\sigma_{\mbox{\rm\tiny diff}}(\Delta)$
is less than or equal to $1$.

{\bf ii)} Let us use the algebra $\Aa_\HV$ common to $\hH$
and $\hHV$ introduced in the appendix. Then both Hamiltonians $H$ and
$(H+V)$ are elements of $\Aa_\HV$ and hence so is $V=(H+V)-H$. Because
$[\XV,\hat{V}]$ is bounded, $V\in\Cc^1(\Aa_\HV)$. Therefore the
diffusion exponent of $\hHV$ is well defined by i). Because it is
defined by means of the invariant ergodic measure $\PP$, Theorem
\ref{theo-stab} implies the result. \hfill $\Box$

\vspace{.2cm}

\noindent {\bf Proof} of Proposition \ref{pro-diffgener}. {\bf i)} Let
$\phi=\sum_{n\in\ZZ^d}\phi_n|n\rangle $ with
$\sum_{n\in\ZZ^d}|\phi_n|<\infty$. By the Schwarz inequality 

$$
\int_{\Omega}d\PP(\omega)
\langle \phi|\delta X^2_{\omega,\Delta}(T)|\phi\rangle 
\;\leq\;
\int_{\Omega}d\PP(\omega)
\langle 0|\delta X^2_{\omega,\Delta}(T)|0\rangle \;\parallel\!\phi
\!\parallel_{\ell^1(\ZZ^d)}^2
\mbox{ , }
$$

\noindent where we have used

\begin{equation}
\XV_\omega(t)-\XV\;=\;U(a)(\XV_{T^a\omega}(t)-\XV)U(a)^*
\label{eq-Xinv}
\end{equation}

\noindent and the invariance of the measure $\PP$. Hence
$\sigma_{\mbox{\rm\tiny diff}}(\Delta)\geq \sigma_{\phi}(\Delta)$.

{\bf ii)} Let $|\psi\rangle
=\Pi_{\omega}({\Delta})|\phi\rangle $ and
$\gamma>2\sigma_{\phi}(\Delta)\geq 0$, then  by Schwarz' inequality

$$
\left(\sqrt{\int^{\infty}_1\frac{dT}{T^{1+\gamma}}\int^T_0\frac{dt}{T}
\langle \psi|\XV_{\omega}(t)^2|\psi\rangle }
-\sqrt{\int^{\infty}_1\frac{dT}{T^{1+\gamma}}
\langle \psi|\XV^2|\psi\rangle }\;\right)^2
\;<\;\infty
\mbox{ . }
$$

\noindent Therefore the exponent defined by using only
$\XV_{\omega}^2(t)$ is necessarily smaller than or equal to
$\sigma_{\phi}(\Delta)$. A similar argument shows the converse
inequality, such that the exponents coincide.
\hfill $\Box$

\vspace{.2cm}

\noindent {\bf Proof} of Theorem \ref{theo-exdiff}. Using DuHamel's formula and
basic properties of projections, one gets

\begin{equation}
\label{eq-condumes}
\int_0^T\frac{dt}{T}\;\TV (|\vec{\nabla}( e^{-\imath Ht})|^2\Pi(\Delta))
\;=\;\int_0^T\frac{dt}{T}\;
 \int_{\Delta\times \RR}
dm(E,E')\;\frac{2-2\cos((E-E')t)}{(E-E')^2}
\mbox{ . }
\end{equation}

\noindent Let now $\gamma\in\RR$ be such that
$2-\gamma>2\sigma_{\mbox{\rm\tiny diff}}(\Delta)$.  Fubini's theorem
then leads to

\begin{eqnarray}
&  & \int_1^{\infty}\frac{dT}{T^{1+(2-\gamma)}}\;\int_0^T\frac{dt}{T}\;
\int_{\Delta\times \RR}
dm(E,E')\;\frac{2-2\cos((E-E')t)}{(E-E')^2} 
\nonumber \\
& &\;\;\;\;\;\;\;\;\;\;\;\;=\;
2\int_{\Delta\times \RR}
dm(E,E')\;|E-E'|^{-\gamma}\int^{\infty}_{|E-E'|}\frac{ds}{s^{1-\gamma}}\;
\frac{s-\sin s}{s^3}
\mbox{ . }
\nonumber
\end{eqnarray}

\noindent The integral over $s$ is bounded for $\gamma\in(0,2)$. 
Therefore, for $\sigma_{\mbox{\rm\tiny diff}}(\Delta)<1$,

\begin{equation}
\label{eq-calcdiff}
2(1-\sigma_{\mbox{\rm\tiny diff}}(\Delta))\;=\;
\sup\;\{\gamma\in\RR\;|\;
\int_{\Delta\times\RR}dm(E,E')\;|E-E'|^{-\gamma}\;<\;\infty\;\}
\mbox{ , }
\nonumber
\end{equation}

\noindent For $\sigma_{\mbox{\rm\tiny diff}}(\Delta)=1$ this is
immediately clear. The  theorem now follows by direct calculation using
Fubini's theorem.
\hfill $\Box$

\vspace{.2cm}

\noindent {\bf Proof} of Theorem \ref{theo-locexpr}. This follows from
similar calculations as in the proof of Theorem \ref{theo-exdiff}
above.
\hfill $\Box$

\vspace{.2cm}

\noindent {\bf Proof} of Theorem \ref{theo-Liou}.  Let us consider
$\imath \Ll_H$ as a self-adjoint operator on $L^2(\Aa,\TV)$. Then its
spectral measure $\rho_{\vec{J}}$ associated to $\vec{J}=\vec{\nabla}
H\in L^2(\Aa,\TV)$ is defined by (\ref{def-Liou}) (the existence is
guaranteed for by the Riesz-Markov theorem). It can be verified by
direct calculation that for any polynomial $f$,

$$
\int d\rho_{\vec{J}}(\epsilon) \;f(\epsilon)\;=\;
\int dm(E,E')\;f(E-E')\mbox{ . }
$$

\noindent By density this extends to any $f\in C_0(\RR)$ such that
the measures coincide. Using this in (\ref{eq-calcdiff}) for the case
$\Delta=\RR$ now allows to conclude. The second statement follows from
the symmetry (\ref{eq-sym}).
\hfill $\Box$

\vspace{.2cm}

\noindent {\bf Proof} of Theorem \ref{theo-exdiff2}.
Fubini's theorem and a contour integration in the upper half plane 
shows in the first place that

$$
2\pi\imath\int_{\RR}da\;S_m(a+\imath {\epsilon},
a-\imath {\epsilon}) 
\;=\;
\int_{\RR^2}dm(E,E')\;\frac{1}{E-E'-2\imath\epsilon}
\mbox{ . }
$$

\noindent The latter is the Green's function
$G_{\rho_{\vec{J}}}(2\imath\epsilon)$ of the measure
${\rho_{\vec{J}}}$. By Theorem \ref{theo-stieltjes}, $\Im
m\;G_{{\rho_{\vec{J}} }}(2\imath\epsilon) \stackrel{{\textstyle
\sim}}{{\scriptscriptstyle \epsilon\downarrow 0}}
\epsilon^{\alpha_{\rho_{\vec{J}}}(0)-1}$. On the other hand,
${\alpha_{\rho_{\vec{J}}}}(0)=2(1-\sigma_{\mbox{\rm\tiny diff}}(\RR))$
by equation (\ref{eq-calcdiff}) for $\Delta=\RR$. The result follows.
\hfill $\Box$

\vspace{.2cm}

\noindent {\bf Proof} of Theorem \ref{theo-Drudeano}. The integrand in
(\ref{eq-kubo3}) is clearly positive such that we may apply Fubini's
theorem. After a change of variables we obtain

$$
\int_1^{\infty}\frac{d\tau_{\mbox{\rm\tiny rel}}}{\tau_{\mbox{\rm\tiny
rel}}^{1+\gamma}}\;
\sigma_{\beta,\mu}(\tau_{\mbox{\rm\tiny rel}},\hat{\omega})
=\frac{q^2}{\hbar}\int_{E\geq E'}dm(E,E')
\frac{f_{\beta,\mu}(E')-f_{\beta,\mu}(E)}{E-E'}
\left(\frac{E-E'}{\hbar}\right)^{\gamma-1}\!\!
\int_{\frac{\hbar}{E-E'}}^{\infty}ds\;\frac{s^{\gamma}}{s^2-1}
\mbox{ .}
$$

\noindent The integral over $s$ is bounded for $-1<\gamma<1$. For
$\beta<\infty$, the only singularity in the integrand of $m$ comes
from the factor $(E-E')^{\gamma-1}$. The result now follows from 
Theorem \ref{theo-exdiff}.
\hfill $\Box$

\vspace{.2cm}

\subsection{The Guarneri bound}
\label{sec-guarneri}

The proof of the following result goes back to Guarneri
\cite{Gua}, with considerable refinements due to \cite{Las,BCM}.
We refer the reader to \cite{Las,BCM} for a proof.

\vspace{.2cm}

\noindent {\bf Theorem} {\sl
Let $H$ be a Hamiltonian on $\HH=\ell^2(\ZZ^d)$ and $\Pi\in\MH$ be its
spectral family. Let $\Delta$ be a Borel spectral subset and
$\phi\in\HH$ satisfying $\langle
\phi|\Pi(\Delta)X^2\Pi(\Delta)|\phi\rangle <\infty$.  If $\rho_{\phi}$
is the spectral measure of $H$ associated to $\phi$ and
$\sigma_{\phi}(\Delta)$ is defined by {\rm Definition
\ref{def-exdiff}} with $\Omega$ reduced to one point, then}

$$
{\alpha}^+_{\rho_{\phi}}(\Delta)\;\leq\;
d\cdot\sigma_{\phi}(\Delta)
\mbox{ . }
$$

\vspace{.2cm}

\begin{rem} {\rm The result can be 
generalized to the study of other moments of the position operator:

$$
\int_0^T\frac{dt}{T}\;\langle \phi|
\Pi(\Delta)\;|X|^{\eta}(t)\;\Pi(\Delta)
|\phi\rangle 
\;=\;
 \sum_{n\in\ZZ^d}\;
|n|^{\eta}\;C_T(\phi,n,\Delta)
\;\stackrel{{\textstyle \sim}}{{\scriptscriptstyle T\uparrow
\infty}} \; T^{\eta\cdot\sigma_{\phi}^{\eta}(\Delta)}
\mbox{ . }
$$

\noindent $\eta=2$ corresponds to  the situation above. The inequality
then reads ${\alpha}^+_{\rho_{\phi}}(\Delta)\leq
d\cdot\sigma_{\phi}^{\eta}(\Delta)$.
}\hfill $\diamond$
\end{rem}

\noindent {\bf Proof} of Theorem \ref{theo-diffgener}{\bf iii)}. Let
$k$ be the smallest integer bigger than $d/2$. Let $\phi\in
\Dd(|\XV|^k)$, the domain of the operator $|\XV|^k$.  If $k=2$, then
the hypothesis $\parallel\!
\pi_{\omega}(\vec{\nabla}^2H)\!\parallel<\infty$ and

$$
|\XV|^2\pi_{\omega}(H)|\phi\rangle\;=\; 
\pi_{\omega}(\vec{\nabla}^2H)|\phi\rangle
+2  \pi_{\omega}(\vec{\nabla}H)\cdot \XV|\phi\rangle+
\pi_{\omega}(H)|\XV|^2|\phi\rangle
$$

\noindent imply that $\pi_{\omega}(H)$ leaves the  domain
$\Dd(|\XV|^2)$ invariant. By functional calculus, $f(\pi_{\omega}(H))$
also leaves  $\Dd(|\XV|^2)$ invariant for any $f\in C^\infty(\RR)$.
Similar arguments treat the case of other $k$'s. We set

$$
\Dd(\omega,\Delta)\;=\;\mbox{span}\{f(\pi_{\omega}(H))|n\rangle\;
|\;n\in\ZZ^d,\;f\in C^\infty(\RR),\;
\mbox{supp}(f)\subset\Delta\}
\mbox{ , }
$$

\noindent where supp$(f)$ denotes the support of $f$.
As $\Delta$ is an open interval, $\Dd(\omega,\Delta)$ is dense in
$\HH(\omega,\Delta)=\Pi_{\omega}(\Delta)\HH$.
Moreover, $\Dd(\omega,\Delta)\subset\Dd(|\XV|^k)$ and

$$
\parallel\! \phi\!\parallel_{\ell^1(\ZZ^d)}\;\leq\;
\left(\sum_{n}\frac{1}{|n|^{2k}}\right)^{\frac{1}{2}}
\;\left(\sum_{n}|n|^{2k}|\phi_n|^2\right)^{\frac{1}{2}}
$$

\noindent shows that $\Dd(\omega,\Delta)\subset\ell^1(\ZZ^d)$.
Finally, let $\Dd'$ be a countable subset of $\Dd(\omega,\Delta)$
still dense in $\HH(\omega,\Delta)$.

\noindent Now for any $\phi\in\Dd'$, Guarneri's inequality given above 
shows that
${\alpha}^+_{\rho_{\omega,\phi}}(\Delta)\;\leq\;
d\cdot\sigma_{\omega,\phi}(\Delta)$. Thus

$$
\sup_{\phi\in\Dd'}
{\alpha}^+_{\rho_{\omega,\phi}}(\Delta)
\;\leq\;d\sup_{\phi\in\Dd'}
\sigma_{\omega,\phi}(\Delta)
\mbox{ . }
$$

\noindent Because of Proposition \ref{prop-Piexpogene2},  the left
hand side is equal to $\alpha^+_{\Pi_\omega}(\Delta)$ and  by Theorem
\ref{theo-dimLDS}  to ${\alpha}^+_{\mbox{\rm\tiny LDOS}}(\Delta)$.
Recall that $\sigma_{\omega,\phi}(\Delta)$ is smaller than or equal to
$\sigma_{\phi}(\Delta)$ $\PP$-almost surely. Because $\Dd'$ is
countable, there exists a set $\Omega_1\subset\Omega$ of full
$\PP$-measure, such that
$\sigma_{\omega,\phi}(\Delta)\leq\sigma_{\phi}(\Delta)$ for all
$\phi\in\Dd'$ and $\omega\in\Omega_1$. Therefore
$\Dd'\subset\ell^1(\ZZ^d)$ gives

$$
{\alpha}^+_{\mbox{\rm\tiny LDOS}}(\Delta)
\;\leq\;d\sup_{\phi\in\Dd'}
\sigma_{\phi}(\Delta)
\;\leq\;d\sup_{\phi\in\ell^1(\ZZ^d)}
\sigma_{\phi}(\Delta)
\mbox{ . }
$$

\noindent But by Proposition \ref{pro-diffgener}ii) the right hand
side is equal to $\sigma_{\mbox{\tiny diff}}(\Delta)$.
\hfill $\Box$

\vspace{.2cm}

\noindent {\bf Proof} of Theorem \ref{theo-locpoint}. {\bf i)} and
{\bf iii)} are already proved in \cite{BES}.  (The definition of
$l^2(\Delta)$ chosen in \cite{BES} was slightly different from
(\ref{def-lo}) if, however, (\ref{def-lo}) is satisfied, it can be
easily shown to be equivalent to the condition in \cite{BES}.) {\bf
ii)} follows directly from the proof of of Theorem
\ref{theo-diffgener}ii). 
\hfill $\Box$

\vspace{.5cm}

\section{Example: Anderson model with free random variables}
\label{chap-Andfree}

As can be seen in equation (\ref{eq-2poin}), one needs to know the
2-particle Green function $G^2$ of a given model in order to calculate
the corresponding conductivity measure. There are not many interesting
models known  in which $G^2$ can be calculated exactly. For Bloch
electrons  one can write out an explicit formula for the conductivity
measure  and then determine the diffusion exponent to be equal to $1$.
On the other hand we already discussed the situation for models with
localization in Section \ref{sec-guarneri}. 

A class of solvable and non-trivial models has been studied
by Wegner \cite{Weg} and Khorunzhy and Pastur \cite{KP}. More
recently, Neu and Speicher \cite{NS} considered a generalization of
these models which will be the starting point here. The Hamiltonian in
these models  is given by the usual Anderson Hamiltonian $H=H_0+H_1$
where

\begin{equation}
\label{eq-HamAnd}
H_0\;=\;\sum_{r\neq s\in\ZZ^d}t_{|r-s|}\;|r\rangle \langle s|
\mbox{ , }
\qquad
H_1\;=\;
\sum_{r\in\ZZ^d}v_r\;|r\rangle \langle r|
\mbox{ , }
\end{equation}

\noindent acting on $\ell^2(\ZZ^d)$, but the on-site disorder
potentials $v_r$ are now supposed to be identically distributed free
random variables in the sense of  Voiculescu \cite{Voi} instead of
independent random variables.  Random variables $X_1,X_2,\ldots$  are
free if  $\EE(P_1(X_{r(1)})P_2(X_{r(2)})\cdots P_m(X_{r(m)}))=0$ for
any set of polynomials $P_k$ satisfying $\EE(P_{k}(X_{r(k)}))=0$,
$k=1,\ldots,m$ whenever $r(k)\neq r(k+1)$, $k=1,\ldots,m-1$. We
suppose that $t_{|n|}\leq|n|^{-d-1-\epsilon}$ for some $\epsilon>0$,
so that $\partial_j(H)$ is bounded for any $j=1,\ldots,d$. As
explained in \cite{Voi,NS},  Wegner's $n$-orbital model in the
integrable limit $n\rightarrow\infty$  \cite{Weg}, the Anderson model
in coherent potential approximation (CPA) and formally also the Lloyd
model \cite{Llo} are special cases of the Anderson model with free
random variables. Strictly speaking this model is not covariant in
the sense of Sections \ref{sec-stab} and \ref{sec-Brillouin}.  The
hull is here a non-commutative manifold. We have not studied the
general formalism in detail hoping that the generalization to this
case is indeed straightforward.

\vspace{.1cm}

As a preamble, let us introduce the main tool for the
addition of free random variables, notably Voiculescu's $R$-transform
\cite{Voi}.  Given a Herglotz function $G(z)$ (namely, $G(z)$
satisfies $\Im m(G(z))\Im m(z)<0$),  its $R$-transform is defined
implicitly by the formula

$$
G(z)\;=\;\frac{1}{z-R(G(z))}\mbox{ . }
$$

\noindent $R(G(z))$ has an analytic continuation to the upper half plane
\cite{NS}. 

\vspace{.1cm}

\noindent {\bf Examples:} {\bf i)} Wigner's semicircle law is given by
the probability measure $\eta$ ($\theta\in\RR$):

$$
d\eta(E)\;=\;\frac{1}{2\pi \theta^2}\;\sqrt{4\theta^2-E^2}\;
\chi(E^2\leq 4 \theta^2)\;dE
\mbox{ . }
$$

\noindent Its Green's function and $R$-transform transform can be calculated
explicitly:

$$
G(z)\;=\;\frac{z-\sqrt{z^2-4\theta^2}}{2\theta^2}\mbox{ , }
\qquad
R(z)\;=\;\theta^2\;z
\mbox{ . }
$$

\noindent Note that $\Im m (R(z)) \leq \theta^2\Im m(z)$ for $\Im
m(z)<0$.

{\bf ii)} Let $\eta$ be the uniform distribution on $[-1,1]$. Then
its Green's function and $R$-transform transform are

$$ 
G(z)\;=\;\frac{1}{2}\mbox{ Log}\frac{z-1}{z+1}\mbox{ , }
\qquad
R(z)\;=\;\frac{1}{\mbox{tanh}z}-\frac{1}{z}\mbox{ . }
$$

\noindent If $z=x-\imath y$, $y>0$, then 

$$
\Im m (R(z))\;=\;\frac{\sin(2y)}{\cos(2y)-\mbox{cosh}(2x)}
-\frac{y}{x^2+y^2}
\mbox{ . }
$$

\noindent Hence $R$ is not a Herglotz function in this example.
\hfill $\diamond$

\vspace{.2cm}

Let us first summarize the main results of \cite{NS}.  The
probability distribution of the $v_r$'s is denoted by $\eta$.  The
space of disorder configurations $\Omega$ is a non-commutative measure
space furnished with an expectation ${\bf E}$ which can be calculated
by free convolution technics \cite{Voi}. Transposing the notations of
Sections \ref{sec-stab} and \ref{sec-Brillouin} to the non-commutative
hull $\Omega$, let $\pi_{\omega}(H)$ be the representation of the
Hamiltonian corresponding to an element  $\omega\in\Omega$. The
following Green's functions are needed $(\Im m\;z> 0)$:

$$
G_0(z)\;=\;\langle 0|\frac{1}{z-H_0}|0\rangle 
\mbox{ , }\qquad
G_1(z)\;=\;\int_\RR d\eta(v_r)\;\frac{1}{z-v_r}
\mbox{ , }
$$

\noindent and $G$ is the diagonal  Green's function of the full
Hamiltonian $H$, that is, of the DOS.  Voiculescu's R-transforms
\cite{Voi} of these functions are denoted by $R_0$ and $R_1$. Finally,
the non-diagonal Green's function $G(r-s,z)$ and its Fourier transform
${\tilde G}(q,z)$ are

\begin{equation}
\label{eq-GreFou}
G(r-s,z) \;=\;
{\bf E}_{\omega}(\langle r|\frac{1}{z-\pi_{\omega}(H)}|s\rangle )
\;=\;
\int_{\BB}\frac{d^dq}{(2\pi)^d}\;e^{\imath q\cdot(r-s)}{\tilde G}(q,z)
\mbox{ , }
\end{equation}

\noindent where $\BB=[-\pi,\pi)^d$ is the Brillouin zone. Let further
$E_0(q)$ be the energy dispersion relation of $H_0$, that is  the
Fourier transform as in (\ref{eq-GreFou}) of the function $t$
determining the kinetic Hamiltonian $H_0$.

\vspace{.4cm}

\noindent {\bf Theorem} (Neu and Speicher \cite{NS}) {\it Consider the
Anderson model with free random variables described by the
Hamiltonian} (\ref{eq-HamAnd}) {\it and suppose the support of the
measure $\eta$ of the $v_r$'s to be compact. Then Green's function
satisfies the following equations

\begin{equation}
\label{eq-1green}
G(z)\;=\;G_0(z-R_1(G(z))\;=\;G_1(z-R_0(G(z))\mbox{ , }
\end{equation}
\noindent and
\begin{equation}
\label{eq-greenfou}
{\tilde G}(q,z)\;=\;\frac{1}{z-E_0(q)-R_1(G(z))}
\mbox{ . }
\end{equation}

\noindent Moreover, the 2-particle Green's function defined in}
(\ref{eq-twopart}) {\it satisfies the identity

\begin{eqnarray}
\label{eq-2poi}
G^2(r,s,s',r',z_1,z_2) & = &
G(r-s,z_1)G(s',r',z_2)\;+\;\frac{R_1(G(z_1))-R_1(G(z_2))}{G(z_1)-G(z_2)}\cdot
\nonumber \\
& & \sum_{t\in\ZZ^d}G(r,t,z_1)\;G^2(t,s,s',t,z_1,z_2)\;G(t,r',z_2)
\mbox{ . }
\end{eqnarray}

\noindent Finally, the solution of the usual Anderson model in CPA are
also given by} (\ref{eq-1green}) {\it and}  (\ref{eq-2poi}).

\vspace{.4cm}

These results allow to calculate the diffusion exponent.

\begin{theo}
\label{theo-freeAnd}
The diffusion exponent  $\sigma_{\mbox{\rm\tiny
diff}}=\sigma_{\mbox{\rm\tiny diff}}(\RR)$ of the Anderson model with
free random variables with compactly supported, absolutely continuous
distribution is bigger than or equal to $1/2$.  The DOS is absolutely
continuous. If moreover, $\Im m (R_1(z)) \leq C\Im m(z)$ for  $\Im
m(z)<0$ and $C\in\RR_+$,  then the diffusion exponent is equal to
$1/2$. In particular, the diffusion exponent of the Wegner $n$-orbital
model in the limit $n\rightarrow \infty$ is equal to $1/2$. 
\end{theo} 

\vspace{.2cm}

\noindent {\bf Proof} of Theorem \ref{theo-freeAnd}. According to a
theorem of Voiculescu \cite{Voi2}, a measure obtained by free
convolution of an absolutely continuous measure with any other measure
is absolutely continuous. The DOS is the free convolution of the
spectral measures of $H_0$ and $H_1$ (see \cite{NS}), such that it is
absolutely continuous because $H_0$ is so.

Let us now calculate the Stieltjes transform $S_m(z_1,z_2)$
of the conductivity measure of the free Anderson model.  Because the
kinetic part in (\ref{eq-HamAnd}) is symmetric,  the matrix elements
$\langle r|\vec{\nabla}(H_0)|s\rangle $ only depend on $r-s$ and
therefore comparing with equation (\ref{eq-2poin})  shows that one
actually needs to calculate the function

$$
\Gg^2(r,s,z_1,z_2)\;=\;
\sum_{t\in\ZZ^d}G^2(r,t-s,t,0,z_1,z_2)
\mbox{ . }
$$

\noindent With the notations

$$
\Gg^1(r,z_1,z_2)\;=\;
\sum_{t\in\ZZ^d}G(r-t,z_1)\;G(t,z_2)\mbox{ , }
\qquad
\Rr(z_1,z_2)\;=\;\frac{R_1(G(z_1))-R_1(G(z_2))}{G(z_1)-G(z_2)}
\mbox{ , }
$$

\noindent and using the fact that $G^2(r,s,s',r',z_1,z_2)$ only
depends on three of its first four entries because the $v_r$ are
identically distributed,  summation of equation (\ref{eq-2poi}) leads
to

$$
\Gg^2(r,s,z_1,z_2)\;=\;\Gg^1(r+s,z_1,z_2)+\Rr(z_1,z_2)
\;\Gg^2(0,s,z_1,z_2)\;\Gg^1(r,z_1,z_2)
\mbox{ . }
$$

\noindent Putting $r=0$ and solving for $\Gg^2(0,s,z_1,z_2)$ one gets

\begin{equation}
\label{eq-sticalc}
\Gg^2(r,s,z_1,z_2)\;=\;\Gg^1(r+s,z_1,z_2)+\frac{\Rr(z_1,z_2)
\;\Gg^1(r,z_1,z_2)\;\Gg^1(-s,z_1,z_2)}{1-\Rr(z_1,z_2)\;\Gg^1(0,z_1,z_2)}
\mbox{ . }
\end{equation}

Remark that the quantity $\Gg^2$ only depends on the
one-particle Green function in this model. Because $E_0(q)$ is an even
function, so is  $\tilde{G}(q,z)$ ({\sl cf.} equation
(\ref{eq-greenfou})) and then

$$
\Gg^1(r,z_1,z_2)\;=\;
\int_{\BB}\frac{d^dq}{(2\pi)^d}\;e^{\imath q\cdot r}{\tilde G}(q,z_1)
\;{\tilde G}(q,z_2)
\mbox{ , }
$$

\noindent implies that $\Gg^1(r,z_1,z_2)$ is also an even function in
its first  variable. Now the matrix elements $\langle
0|\vec{\nabla}(H_0)|s\rangle $ change sign as  the sign of $s$ is
changed, and  therefore the second term in (\ref{eq-sticalc}) gives no
contribution to the Stieltjes transform of the conductivity measure in
(\ref{eq-2poin}). Passing to Fourier space the operator
$\vec{\nabla}(H_0)$ becomes multiplication by the gradient of $E_0(q)$
with respect to quasi-moments and subsequent summation in
(\ref{eq-2poin}) leads to the result

\begin{equation}
\label{eq-sticond}
S_m(z_1,z_2)\;=\; 
\int_{\BB}\frac{d^dq}{(2\pi)^d}\;|\vec{\nabla}_q\;E_0(q)|^2\; 
\frac{1}{z_1-E_0(q)-R_1(G(z_1))}\;\frac{1}{z_2-E_0(q)-R_1(G(z_2))}
\mbox{ . }
\end{equation}

For Wegner's $n$-orbital model, this equation has already
been derived by Khorunzhy and Pastur \cite{KP}. Owing to Theorem
\ref{theo-exdiff2} and  equation (\ref{eq-sticond}), it is now
necessary to consider the integral

$$
\int_{\RR}da
\int_{\BB}\frac{d^dq}{(2\pi)^d}\;|\vec{\nabla}_q\;E_0(q)|^2\; 
\frac{1}{|a+\imath\epsilon-E_0(q)-R_1(G(a+\imath\epsilon))|^2}
\mbox{ , }
$$

\noindent and to study its behavior in the limit $\epsilon\rightarrow
0$.  Clearly the integral is strictly bigger than zero because of
contributions for big $|a|$ and therefore the diffusion exponent is
bigger than or equal to $1/2$ according to Theorem \ref{theo-exdiff2}. 

In order to show that the above integral is bounded let us
use its upper bound

\begin{eqnarray}
& & \int_{\RR}da
\int_{\BB}\frac{d^dq}{(2\pi)^d}\; 
\frac{1}{|a+\imath\epsilon-E_0(q)-R_1(G(a+\imath\epsilon))|^2}
\nonumber \\
&  & \;\;\;\;\;\;\;\;=\;\int_{\RR}da \;\int_{\RR}d\Nn_0(E)
\frac{1}{|a+\imath\epsilon-E-R_1(G(a+\imath\epsilon))|^2}
\nonumber \\
&  & \;\;\;\;\;\;\;\;=\; \int_{\RR}da 
\frac{\Im
m\;G(a+\imath\epsilon)}{\Im m(R_1(G(a+\imath\epsilon)))-\epsilon}
\mbox{ , }
\label{eq-lasthelp}
\end{eqnarray}

\noindent where the last equality follows by direct calculation using
the identity (\ref{eq-1green}). The integrand in (\ref{eq-lasthelp})
is bounded by the hypothesis made on $R_1$ and it falls off as $1/a^2$
at infinity. Therefore the integral is bounded and the diffusion
exponent is equal to $1/2$.

In Wegner's $n$-orbital model the $v_r$ are given by
$n\times n$ hermitian random matrices with independent Gaussian
entries.  The model then becomes exactly solvable in the limit
$n\rightarrow \infty$. In fact, Voiculescu has shown that  Gaussian
random matrices are asymptotically free such that the $v_r$ in
(\ref{eq-HamAnd}) are free in the limit $n\rightarrow \infty$. The
distribution of the $v_r$ is then given by Wigner's semicircle law and
thus the hypothesis on $R_1$ is satisfied ({\sl cf.} the above
example).
\hfill $\Box$

\vspace{0.2cm}

\begin{rem} {\rm Let us investigate the Lloyd model. The $v_r$ then
follow the Cauchy distribution with parameter $\gamma>0$. Because all
moments diverge, the calculations in \cite{NS} for the 2-point Green's
function are only formal. Actually, the formal calculations lead to
erroneous results as the following arguments show. One has
$G_1(z)=1/(z+\imath\gamma\;$sign$(\Im m z)) $ and this  implies that
$R_1(z)= \imath\gamma\;$sign$(\Im m z) $. Use this for the calculation
of $S_m$ in  (\ref{eq-sticond}) which is supposed to hold.  Because
the Green's function is a Herglotz function, a contours integration
leads to 

$$
\Re e \left(\int_{\RR}da\;S_m(a+\imath {\epsilon},
a-\imath {\epsilon})\right)\;
=\;\frac{-1}{4\pi}\;\frac{1}{\epsilon+\gamma}
\;\int_{\BB}\frac{d^dq}{(2\pi)^d}\;|\vec{\nabla}_q\;E_0(q)|^2
\mbox{ . }
$$

According to  Theorem \ref{theo-exdiff2} the diffusion
exponent would therefore be equal to $1/2$. This would be independent
of the dimension of the physical space and the strength of the
coupling. In particular, it would imply regular diffusion in the
one-dimensional Lloyd model. But this is in contradiction with the
rigorous results of Aizenman and Molchanov \cite{AiM} which imply that
the diffusion exponent is equal to  zero in this situation as is shown
in \cite{BES}.

It is possible, however, to calculate the exponent of the
DOS of the Lloyd model using equation (\ref{eq-1green}) and Theorem
\ref{theo-stieltjes}. These are then shown to be equal to $1$ which is
a  well known result \cite{CFKS}. In fact, it is shown in \cite{BV}
that the calculations in \cite{NS} are justified for the 1-point
Green's function.
\hfill $\diamond$
}
\end{rem}

\begin{rem} {\rm  Formula (\ref{eq-sticond})  also holds for the case
where $v_r$'s are all equal to $0$. Then $R_1(z)=0$. Therefore a
contour integration leads to

$$
\Re e \left(\int_{\RR}da\;S_m(a+\imath {\epsilon},
a-\imath {\epsilon})\right)\;
=\;\frac{1}{2\pi\imath}
\int_{\BB}\frac{d^dq}{(2\pi)^d}\;|\vec{\nabla}_q\;E_0(q)|^2\;
\frac{1}{2\imath\epsilon}\;
\stackrel{{\textstyle \sim}}{{\scriptscriptstyle \epsilon\downarrow
0}}
\;\epsilon^{-1}
\mbox{ . }
$$

\noindent Comparison with Theorem \ref{theo-exdiff2} shows that
$\sigma_{\mbox{\rm\tiny diff}}(\RR)=1$ as expected for free particles. 
\hfill $\diamond$
}
\end{rem}

\begin{rem} {\rm The two other integrable models  considered by
Khorunzhy and Pastur  \cite{KP} have a conductivity measure which is
absolutely continuous with respect to the Lebesgue measure on $\RR^2$
\cite[eq. (2.29)]{KP}. For such an absolutely continuous conductivity
measure, Theorem \ref{theo-exdiff} implies directly a diffusion
exponent $1/2$.
\hfill $\diamond$
}
\end{rem}

\vspace{.2cm}

\section{Appendix}

Let us consider a topological dynamical system $(\Omega,T,\ZZ^d)$
where $\Omega$ is a compact metrisable space and $T$ is an action of
$\ZZ^d$ by homeomorphisms. A point $\omega\in\Omega$ is called
wandering  whenever there is an open neighborhood $\Uu$ such that
$T^a(\Uu)\cap\Uu=\emptyset$ for any
$a\in\ZZ^d_*=\ZZ^d\backslash\{0\}$. The set  $\Omega^{\mbox{\rm\tiny
w}}$ of all wandering points is open. Hence its complement
$\Omega^{\mbox{\rm\tiny nw}}$, the set of all non-wandering points, is
compact.

\begin{proposi}
\label{prop-nonwandering}
Any $T$-invariant probability $\PP$ is supported on 
$\Omega^{\mbox{\rm\tiny nw}}$.
\end{proposi}

\noindent {\bf Proof.} Let $\omega\in\Omega^{\mbox{\rm\tiny w}}$ and
$\Uu$ be a neighborhood satisfying $T^a(\Uu)\cap\Uu=\emptyset$ for any
$a\in\ZZ^d_*$. Then $T^a(\Uu)\cap T^b(\Uu)=\emptyset$ for any $a\neq
b\in\ZZ^d$. The $\sigma$-additivity and $T$-invariance of $\PP$ imply

$$
\PP(\Omega)\;\geq\;\PP(\bigcup_{a\in\ZZ^d}T^a(\Uu))\;=\;
\sum_{a\in\ZZ^d}\PP(\Uu)
\mbox{ , }
$$

\noindent so that $\PP(\Uu)=0$. Therefore the support of $\PP$ does
not contain $\Omega^{\mbox{\rm\tiny w}}$.
\hfill $\Box$

\vspace{.2cm}

Let us now consider the non-wandering points of the hull
$(\Omega_\hH,T,\ZZ^d)$ associated to a homogeneous Hamiltonian $\hH$
on $\Hh=\ell^2(\ZZ^d)$ by (\ref{eq-hull}). We denote $\hH\in\Omega$ by
$\omega_0$. The orbit of $\omega_0$ is the set
$\mbox{Orb}(\omega_0)=\{T^a\omega_0|a\in\ZZ^d\}$.

\begin{lemma}
\label{lem-nonwandering} Let us introduce the set

$$
\Omega_\hH^\infty\;=\;\{\omega\in\Omega\;|\;\exists
\;\,\mbox{\rm sequence }(a_n)_{n\in\NN} \;\mbox{\rm in }
\ZZ^d,\;\lim_{n\rightarrow\infty}|a_n|=\infty,\;\mbox{\rm s.t. }
\lim_{n\rightarrow\infty}T^{a_n}\omega_0=\omega\,\}
\mbox{ . }
$$

\noindent If $\omega_0$ is not an element of $\Omega_\hH^\infty$, then
$\Omega_\hH^{\mbox{\rm\tiny w}}=\mbox{\rm Orb}(\omega_0)$. If
$\omega_0\in\Omega_\hH^\infty$, then $\Omega_\hH^{\mbox{\rm\tiny w}}=
\emptyset$. In both cases, $\Omega_\hH^{\mbox{\rm\tiny nw}}=
\Omega_\hH^\infty$.
\end{lemma}

\noindent {\bf Proof.} Let $\omega$ belong to $\Omega$. Then there
exists a sequence $(a_n)_{n\in\NN}$, $a_n\in\ZZ^d$, such that
$\lim_{n\rightarrow\infty}T^{a_n}\omega_0=\omega$. Using the one-point
compactification $\ZZ^d\cup\{\infty\}$ of $\ZZ^d$, one can extract a
convergent subsequence $(a_{n_k})_{k\in\NN}$ with
$\lim_{k\rightarrow\infty}T^{a_{n_k}}\omega_0=\omega$. If
$\lim_{k\rightarrow\infty}a_{n_k}=\infty$, then
$\omega\in\Omega_\hH^\infty$, otherwise $(a_{n_k})$ is a stationary
sequence on some $a\in\ZZ^d$ and
$\omega=T^a\omega_0\in\mbox{Orb}(\omega_0)$. Hence $\Omega=
\Omega_\hH^\infty\cup\mbox{Orb}(\omega_0)$, but this decompostion is
not necessarily disjoint.

We next show that
$\Omega_\hH^\infty\subset\Omega_\hH^{\mbox{\rm\tiny nw}}$. For
$\omega\in\Omega_\hH^\infty$ there exists an unbounded sequence
$(a_n)_{n\in\NN}$ in $\ZZ^d$ such that
$\lim_{n\rightarrow\infty}T^{a_n}\omega_0=\omega$. Hence for any
neighborhood $\Uu$ of $\omega$ there is a $N\in\NN$ such that
$\{T^{a_n}\omega_0\,|\,n\geq N\}\subset\Uu$. Choose $a_n\neq a_m$,
$n,m\geq N$. Then $T^{a_n}\omega_0\in\Uu$ and
$T^{a_m-a_n}(T^{a_n}\omega_0)\in\Uu$ so that
$T^{a_m-a_n}(\Uu)\cap\Uu\neq\emptyset$. As this holds for any
neighborhood of $\omega$, $\omega\in\Omega_\hH^{\mbox{\rm\tiny nw}}$.

Now either $\omega_0\in\Omega_\hH^{\mbox{\rm\tiny w}}$ or
$\omega_0\in\Omega_\hH^{\mbox{\rm\tiny nw}}$. In the first case,
$\mbox{Orb}(\omega_0)\subset\Omega_\hH^{\mbox{\rm\tiny w}}$. According
to the above, $\mbox{Orb}(\omega_0)=\Omega_\hH^{\mbox{\rm\tiny w}}$
and $\Omega_\hH^{\mbox{\rm\tiny nw}}=\Omega_\hH^\infty$. To deal with
the second case, let us choose a metric on $\Omega$ compatible with
the topology of $\Omega$. Open balls of radius $r$ around
$\omega\in\Omega$ are denoted by $B(\omega,r)$. Because
$\omega_0\in\Omega_\hH^{\mbox{\rm\tiny nw}}$, there exists for any
$k\in\NN$ an $a_k\in\ZZ^d_*$ such that  $B(\omega_0,\frac{1}{k})\cap
T^{a_k}B(\omega_0,\frac{1}{k})\neq\emptyset$. Therefore the sequence
$T^{a_k}\omega_0$ converges to $\omega_0$. As $a_k\neq 0$ for all $k$,
$\omega\in\Omega_\hH^\infty$. Hence
$\mbox{Orb}(\omega_0)\subset\Omega_\hH^\infty$ and
$\Omega_\hH^{\mbox{\rm\tiny nw}}=\Omega_\hH^\infty=\Omega$.
\hfill $\Box$

\vspace{.2cm}

\noindent {\bf Proof} of Theorem \ref{theo-stab}.  Let
$\hat{V}=\hat{V}^*$ be a compact operator of $\Hh=\ell^2(\ZZ^d)$. The
hulls of the Hamiltonian $\hH$ and the perturbed Hamiltonian $\hHV$
are denoted by $\Omega_\hH$ and $\Omega_\hHV$. $\hat{V}$ compact
implies that s-$\lim_{n\rightarrow\infty}U(a_n)\hat{V}U(a_n)^*=0$ for
an unbounded sequence $|a_n|\rightarrow\infty$. Hence Lemma
\ref{lem-nonwandering} implies $\Omega_\hH^{\mbox{\rm\tiny nw}}
=\Omega_\hHV^{\mbox{\rm\tiny nw}}$. Proposition
\ref{prop-nonwandering} and Lemma \ref{lem-nonwandering} show the
other claims.
\hfill $\Box$

\vspace{.2cm}

For a compact perturbation $\hat{V}$, let us introduce the
common hull of $\hH$ and $\hHV$ by

$$
\Omega_\HV\;=\;
\overline{\{(U(a)\hH U(a)^{-1},U(a)\hHV U(a)^{-1})\;|\;a\in\ZZ^d\} 
}^{\mbox{\rm\tiny s}}
\mbox{ , }
$$

\noindent where the closure is taken with respect to the strong
topology on the Cartesian product $\Bb(\Hh)\times\Bb(\Hh)$. By the
arguments of the proof above, $\Omega_\HV^{\mbox{\rm\tiny
nw}}=\{(\omega,\omega)\,|\, \omega\in\Omega_\hH^{\mbox{\rm\tiny
nw}}\}$. Hence $(\Omega_\HV^{\mbox{\rm\tiny nw}},T,\ZZ^d)$ and
$(\Omega_\hH^{\mbox{\rm\tiny nw}},T,\ZZ^d)$ are homeomorhpic as
dynamical systems. Denoting by $p_\hH$ ($p_\hHV$) the projection from
$\Omega_\HV$ to its first (second) component, we have the following
picture:

$$
\begin{array}{ccccc}
\Omega_\hH & 
\stackrel{{\longleftarrow}}{{\scriptscriptstyle \;\;\;p_\hH}} &
\Omega_\HV &
\stackrel{{\longrightarrow}}{{\scriptscriptstyle p_\hHV}} &
\Omega_\hHV
\\
\cup & & \cup & & \cup \\
\Omega_\hH^{\mbox{\rm\tiny nw}} & 
\stackrel{{\longleftrightarrow}}{{\scriptscriptstyle\;\;p_\hH}} &
\Omega_\HV^{\mbox{\rm\tiny nw}} &
\stackrel{{\longleftrightarrow}}{{\scriptscriptstyle p_\hHV}} &
\Omega_\hHV^{\mbox{\rm\tiny nw}}
\end{array}
$$

To conclude, let us consider the crossed product
$C^*$-algebras  $\Aa_\hH$,  $\Aa_\hHV$ and $\Aa_\HV$ constructed from
the dynamical systems on $\Omega_\hH$, $\Omega_\hHV$ and $\Omega_\HV$
by the procedure of Section \ref{sec-stab}. There is an injection
$i_\hH:\Aa_\hH \hookrightarrow\Aa_\HV$ which sends $f\in
C_0(\Omega_\hH\times\ZZ^d)$ to a function $i_\hH(f)\in
C_0(\Omega_\HV\times\ZZ^d)$ independent of the second argument in
$\Omega_\HV$. Similarly one has
$i_\hHV:\Aa_\hHV\hookrightarrow\Aa_\HV$. On the other hand, because
the dynamical systems $(\Omega_\HV^{\mbox{\rm\tiny nw}},T,\ZZ^d)$ and
$(\Omega_\hH^{\mbox{\rm\tiny nw}},T,\ZZ^d)$ are homeomorphic, they
only give rise to one crossed product $\Aa_\hH^{\mbox{\rm\tiny nw}}$.
Let us introduce the ideal

$$
J_\hH^{\mbox{\rm\tiny nw}}\;=\;\{A\in\Aa_\hH\;|\;\pi_\omega(A)=0
\;\;\forall\;\omega\in\Omega_\hH^{\mbox{\rm\tiny nw}}\}
\mbox{ , }
$$

\noindent and similarly $J_\hHV^{\mbox{\rm\tiny nw}}$ and
$J_\HV^{\mbox{\rm\tiny nw}}$. Then $\Aa_\hH^{\mbox{\rm\tiny nw}}$ is
the quotient of $\Aa_\hH$ (or $\Aa_\hHV$ or $\Aa_\HV$) with respect to
$J_\hH^{\mbox{\rm\tiny nw}}$ (or $J_\hHV^{\mbox{\rm\tiny nw}}$ or
$J_\HV^{\mbox{\rm\tiny nw}}$). Hence by using the surjective
applications on the quotient, one has

$$
\begin{array}{ccccc}
\Aa_\hH & 
\stackrel{{\longrightarrow}}{{\scriptscriptstyle \;\;i_\hH}} &
\Aa_\HV &
\stackrel{{\longleftarrow}}{{\scriptscriptstyle i_\hHV}} &
\Aa_\hHV
\\
 & \searrow & \downarrow & \swarrow &  \\
 &  & \Aa_\hH^{\mbox{\rm\tiny nw}} & &
\end{array}
\nonumber
$$

Let now a trace $\TV$ be defined on $\Aa_\hH$ (or $\Aa_\hHV$
or $\Aa_\HV$) by (\ref{eq-tracepervol}). As $\TV$ vanishes on the
ideals $J_\hH^{\mbox{\rm\tiny nw}}$ (or $J_\hHV^{\mbox{\rm\tiny nw}}$
or $J_\HV^{\mbox{\rm\tiny nw}}$) by Proposition
\ref{prop-nonwandering}, it is well defined on
$\Aa_\hH^{\mbox{\rm\tiny nw}}$.

\end{document}